\documentclass[10pt]{article}
\title{The largest eigenvalues of sample covariance matrices for a spiked population:\\ diagonal case.}
\author{Delphine F\'eral \thanks{Institut de Math\'ematiques de
    Bordeaux, Universit\'e  Bordeaux 1, 351 Cours de la Lib\'eration,
    F-33405 Talence Cedex. E-mail: delphine.feral@math.u-bordeaux1.fr}
  \, and Sandrine P\'ech\'e \thanks{Institut Fourier, 100 Rue des Maths,
    BP 74, F-38402 St Martin d'Heres. E-mail: Sandrine.Peche@ujf-grenoble.fr}}
\usepackage{epsfig}
\usepackage{amsfonts}
\usepackage{amssymb}
\usepackage{amsthm}
\usepackage{amstext}
\usepackage{amscd}
\usepackage{amsmath}
\usepackage{delarray}
\begin{document}
\maketitle
\newtheorem{theo}{Theorem}[section]
\newtheorem{prop}{Proposition}[section]
\newtheorem{lemme}{Lemma}[section]
\newtheorem{conjecture}{Conjecture}[section]
\newtheorem{definition}{Definition}[section]
\newtheorem{fact}{Fact}[section]
\newtheorem{hyp}{Assumption}[section]
\newtheorem{com}{Comments}[section]
\theoremstyle{remark}
\newtheorem{rem}{Remark}
\newtheorem{remark}{Remark}[section]
\newtheorem{Remark}{Remark}[section]
\newtheorem*{NotRem}{Notational Remark}
\newcommand{\bremnot}{\begin{Notational Remark}}
\newcommand{\eremnot}{\end{Notational Remark}}
\newcommand{\brem}{\begin{remark}}
\newcommand{\erem}{\end{remark}}
\newcommand{\bconj}{\begin{conjecture}}
\newcommand{\econj}{\end{conjecture}}
\newcommand{\bdefi}{\begin{definition}}
\newcommand{\edefi}{\end{definition}}
\newcommand{\bt}{\begin{theo}}
\newcommand{\bfa}{\begin{fact}}
\newcommand{\efa}{\end{fact}}
\newcommand{\Si}{\Sigma}
\newcommand{\et}{\end{theo}}
\newcommand{\bp}{\begin{prop}}
\newcommand{\ep}{\end{prop}}
\newcommand{\bl}{\begin{lemme}}
\newcommand{\el}{\end{lemme}}
\newcommand{\be}{\begin{equation}}
\newcommand{\mbE}{\mathbb{E}}
\newcommand{\tTr}{{\rm{Tr}}}
\newcommand{\ee}{\end{equation}}
\newcolumntype{L}{>{$}l<{$}}
\newcommand{\mbC}{\mathbb{C}}
\newcommand{\mbP}{\mathbb{P}}

\begin{abstract}
We consider large complex random sample covariance matrices obtained
from ``spiked populations'', that is when the true covariance matrix
is diagonal with all but finitely many eigenvalues equal to one. We
investigate the limiting behavior of the largest eigenvalues when the
population and the sample sizes both become large. Under some
conditions on moments of the sample distribution, we prove that the asymptotic fluctuations 
of the largest eigenvalues are the same as for a complex Gaussian
sample with the same true covariance. The real setting is also considered. 
\end{abstract}

\section{Introduction and results}
Sample covariance matrices are fundamental to multivariate statistics. Their spectral properties are e.g. important for Principal Component Analysis. In the case where the population size remains ``small'' while the sample size becomes sufficiently large, these spectral properties are well-understood. It is a classical probability result that the sample covariance matrix is a good approximate of the population covariance. Nowadays it is of strong interest to study the case where both the sample and population sizes become large, due to the large amount of data available.
In this setting, the study of asymptotic spectral properties of sample covariance matrices has many applications. The behavior of Principal Component Analysis has first to be understood. We refer the reader to \cite{Johnstone}
and \cite{NEKlarginterest} for a review of other statistical applications.
Other examples of applications include genetics \cite{Paterson},
mathematical finance \cite{PlerousGRAGS}, \cite{Bouchaud}, \cite{LalouxCPB},
\cite{MalevergneS}, wireless communication \cite{Telatar}, physics
of mixture \cite{SearC} and statistical
learning \cite{HoyleR}.\\
In this paper, we investigate the limiting distribution of the largest
eigenvalues of sample covariance matrices for some so-called ``spiked
population models''. Such models have been introduced for a Gaussian sample in
\cite{Johnstone} and correspond to the case where the true covariance
is a small rank perturbation of the Identity matrix. In this paper, both the impact of the largest eigenvalues of the true covariance and that of the distribution of the sample on the asymptotic behavior of
the largest eigenvalues are investigated. The fluctuations of the largest eigenvalues of some non necessarily Gaussian samples are compared to those of a
Gaussian sample with the same true covariance. These questions are mainly motivated by statistical applications. Indeed
some statistical tests are based on the conjecture that the behavior of the largest eigenvalues of spiked sample covariance matrices is the same as for a Gaussian sample provided the sample distribution
is close to a Gaussian distribution (see e.g. \cite{Paterson}).

\subsection{Model and results} \label{sec: model}
Let $X=X_N$ be a $N \times p$ complex (resp. real) random matrix
such that $\{ \Re e X_{ij}, \, \Im m  X_{ij}; \, 1 \leq i \leq N,
1 \leq j \leq p \}$ (resp. $\{ X_{ij}, \, 1 \leq i \leq
N, 1 \leq j \leq p \}$) are real independent random variables satisfying for all $1 \leq i \leq N$ and $1 \leq j \leq p$:
\begin{itemize} 
\item[${\rm(H_1)}$] $\mathbb{E}X_{ij}=0 \, \text { and } \, \mbE  \left (\Re e
X_{ij}\right) ^2=\mbE  \left (\Im m X_{ij}\right)^2= \sigma^2/2 \text{ (resp. $\mbE
X_{ij}^2=\sigma^2$)}$;
\item[${\rm(H_2)}$] there exists a constant $ C_o>0$ independent of $N,p $ (and $(i,j)$) such that \\
$\forall
k>0, \quad \mbE|X_{ij}|^{2k}\leq \left (C_ok\right)^k;$
\item[${\rm(H_3)}$] $\mbE \left ((\Re e X_{ij})^{2k+1}\right ) =\mbE \left ((\Im m X_{ij})^{2k+1}\right ) =0$ (resp. $ \mbE \left (X_{ij}^{2k+1}\right ) =0$) $\forall k\geq 0$.
\end{itemize}
To avoid technicalities, we assume
throughout the paper that $p\geq N$. Here the size of the matrix $X$
goes to infinity in such a way that if we set $\gamma_N=p/N$,
\begin{equation}\label{def: gamma} \exists \gamma\geq 1 \text{ such
that }\gamma_N\to \gamma \text{ as }N \to \infty.
\end{equation}
Let $r$ be a given integer independent of $N$ and $p.$ Let also
$\pi_1 \geq \pi_2 \geq \ldots \geq \pi_r >1$ be given real numbers,
all of which are independent of $N$ and $p$. The covariance matrix
$\Sigma=\Sigma _N$ is the $N \times N$ diagonal matrix \be\Sigma=
\text{diag}(\pi_1,\pi_2, \ldots, \pi_r,1, \ldots,1).\label{def:
Si}\ee

The goal of this paper is to describe the large-$N$-limiting
distribution of the largest
eigenvalues of the spiked model defined by \be \label{def:
VN}V_N:=\frac{1}{p}\Sigma^{1/2}XX^* \Sigma^{1/2}.\ee 
Note that the spectral properties of the associated matrix
$V'_N=\frac{1}{p}X^*\Sigma X$ can be deduced from those of
$V_N$ since their non-zero eigenvalues are equal.\\
Throughout this paper, the white (or null) model corresponds to
$\Sigma=Id$ (or $r=0$) and is called \be \label{def: MN}
M_N:=\frac{1}{p}XX^*.\ee When the entries of $X$ are further assumed
to be Gaussian random variables, we write $M_N^G$ (resp. $V_N^G$) instead of
$M_N$ (resp. $V_N$). $M_N^G$ is then a matrix from the
so-called Laguerre unitary (resp. orthogonal) ensemble (LUE (resp. LOE)) of
parameter $\sigma$, also known as the complex (resp. real) Wishart ensembles.

\paragraph{}
First, let us consider the global behavior of the spectrum. Let
$\lambda_1(V_N) \geq \lambda_2 (V_N) \geq \cdots \geq \lambda_N
(V_N)$ be the ordered eigenvalues of $V_N$ and let $\mu_N$ be the
spectral measure defined by $\mu_N =\frac{1}{N}\sum_{i=1}^N
\delta_{\lambda_i(V_N)}$. Setting 
\begin{equation}\label{Defu+}
u_{\pm}=\sigma^2(1\pm \gamma^{-1/2})^2,
\end{equation}
it is a well-known result of \cite{MP} that, for any matrix $\Sigma$ given by (\ref{def:
Si}) (including $\Sigma=Id$), $\mu_N$ a.s. converges as
$N \to \infty$ to the Marchenko-Pastur
distribution $\mu_{MP}$ whose density is $
\frac{d\mu_{MP}(x)}{dx}=\frac{\gamma}{2\pi x \sigma^2}
\sqrt{(u_+-x)(x-u_-)}1_{[u_-, u_+]}(x).$  The global behavior of
the spectrum is thus not impacted by the spiked structure of $\Sigma$. 
\paragraph{}The situation is drastically different for the largest eigenvalues.
Let us first recall the well-known
asymptotic behavior of the largest eigenvalues of $M_N$. This asymptotic behavior has been identified for the complex or real Wishart
ensembles $M_N^G$ in \cite{Johansson} and \cite{Johnstone} and later extended to a much wider class of white sample covariance matrices $M_N$ in
\cite{PecheWishart}. To be more precise, we need the following
definitions. We denote by $Ai$ the standard Airy function. Define
the Airy kernel by
${\rm
A}(u,v)=\frac{Ai(u)Ai'(v)-Ai'(u)Ai(v)}{u-v}$ and let ${\rm A}_x$ be
the operator acting on $L^2((x, +\infty))$ with kernel ${\rm A}(u,
v)$. Let $F_{\rm GU(O)E}$ be the GU(O)E Tracy-Widom distribution defined in \cite{TWAi},
which is the limiting distribution of the 
largest eigenvalue of 
the Gaussian unitary (resp. orthogonal) ensemble (GU(O)E) as the
size tends to infinity. It can in particular be shown that $F_{\rm
GUE}$ is given by the Fredholm determinant $F_{\rm
GUE}(x)=\det(1-{\rm A}_x)$. More generally, given an integer $K
\geq 1$, we denote by $F^{K}_{\rm GU(O)E}$ the limiting joint
distribution of the $K$ largest eigenvalues of the GU(O)E (the
precise definitions are given in \cite{TWAi} and \cite{TWAi2}).
Last we define $$
\rho_{N}=\sigma^2\left ( 1+\gamma_N^{-1/2}\right)^2 \quad \text{ and
} \quad \sigma_{N}=\gamma_N^{-1/2}\sigma^2\left ( 1+\gamma_N^{-1/2}\right)^{4/3}.$$
The next theorem has been proved in the more general case where $\gamma \in [0, \infty]$.

\bt \label{theo: fluctuwhite} \cite{Johansson}, \cite{Johnstone},
\cite{PecheWishart} Let $K \geq 1$ be an integer. Let $M_N$ be
a complex (or real) random matrix given by (\ref{def: MN}) and
assume that $X$ satisfies ${\rm{(H_1)-(H_3)}}$. Then, for any $(x_1, \ldots, x_K)
\in \mathbb R ^K$, one has that
$$\lim_{N \to \infty}\mbP \left ( \frac{N^{\frac{2}{3}}}{\sigma_{N}} (\lambda_i(M_N) -\rho_{N} )\leq x_i, \:
\forall i=1, \ldots, K \right) = F^K_{\rm GU(O)E}(x_1, x_2, \ldots,
x_K).$$ \et

Theorem \ref{theo: fluctuwhite} implies that a.s. $\lim_{N \to
  \infty}\lambda_1(M_N)=u_+$ (this result is proved in a more general
setting in \cite{Yin}). 
The situation may be quite different if $\Sigma$ is chosen as in
(\ref{def: Si}). The first results in this direction have been
obtained in \cite{BaikGBAPeche} for the complex Gaussian spiked
population model $V_N^G$. Therein the authors point
out a phase transition phenomenon for the fluctuations of the largest eigenvalue according
to the value of the largest eigenvalue(s) of the covariance matrix
$\Sigma$. To state the result, further definitions are needed. For
all $m \geq 1$, set
$$
s^{(m)}(u)=\int_{\infty e^{\frac{5i\pi}{6}}}^{\infty
e^{\frac{i\pi}{6}}}
\frac{e^{\{iua+\frac{ia^3}{3}\}}}{2\pi}\frac{da}{(ia)^m} 
\text{, } t^{(m)}(v)=\int_{\infty
e^{\frac{5i\pi}{6}}}^{\infty e^{\frac{i\pi}{6}}}
\frac{e^{\{iva+\frac{ia^3}{3}\}}}{2\pi}(ia)^{m-1}da,
$$
which are integrals or derivatives of the standard Airy function
(cf. \cite{BaikGBAPeche}). For any $k \geq 1$, define the
distribution function $F_k$ (see \cite{Baik}) by \be
F_{k}(x)=\det(1-A_x)\det\left(\delta_{m,n}-<\frac{1}{1-A_x}s^{(m)},
t^{(n)}>\right)_{1\leq m,n\leq k}, \ee for any real $x$, where
$<.,.>$ denotes the (real) inner product of functions in $L^2((x, \infty))$. 
Define also 
\begin{eqnarray}&& w_c:=1+\frac{1}{\sqrt \gamma} ,
\label{wcrit}	\\
&& \tau(\pi _1)= \sigma ^2 \, \pi_1 \left ( 1+\frac{\gamma ^{-1 }}{\pi_1-1}
\right ), \quad \sigma(\pi
_1)= \sigma ^2 \,\pi_1\sqrt{ 1-\gamma^{-1}/{(\pi_1-1)^2}}.\label{tausigmapi1}
\end{eqnarray}
Note that if $\pi _1 > w_c$ (resp. $\pi _1 = w_c$, resp. $\pi _1
<w_c$), then $\tau(\pi _1)> u_+$ (resp. $\tau(\pi _1)= u_+$, resp.
$\tau(\pi _1)< u_+$).
\paragraph{}
We here give only the asymptotic behavior of the largest eigenvalue of the complex non-white Wishart ensemble (see Remark \ref{rem ExtK} for some extentions).

\bt \cite{BaikGBAPeche} \label{theo: BaikGBAPeche} 
Consider the sequence of complex Wishart matrices $(V_N^G)$ when $\Sigma$ is given by (\ref{def: Si}). Let $1 \leq k \leq r$ be an integer. For any real $x$, one has that
\begin{itemize}
\item[(i)] If $\pi_1=\ldots=\pi_k>w_c$ and $\pi_{k+1}<\pi_1$ then
$$\lim_{N,p\rightarrow \infty}\mathbb{P}\Bigl
(\frac{\sqrt N}{\sigma (\pi _1) }\left ( \lambda_1(V_N^G)-\tau
(\pi_1)\right )\leq x\Bigr)=G_k(x),$$ where $G_k$ is the
distribution of the largest eigenvalue of the un-normalized GUE
random matrix $H=(H_{ij})_{i,j=1}^k$ with i.i.d. complex standard
Gaussian entries above the diagonal.
\item[(ii)] If $\pi_1=\ldots=\pi_k=w_c$ and $\pi_{k+1}<w_c$ then $$\lim_{N,p\rightarrow \infty}\mathbb{P}\Bigl
(\frac{N ^{2/3}}{\sigma_{N}} \,  ( \lambda_1(V_N^G)-\rho_{N} ) \leq
x\Bigr)=F_{k}(x).$$
\item[(iii)] If $\pi_1<w_c$ then
$\displaystyle{\lim_{N,p\rightarrow \infty}\mathbb{P}\Bigl (\frac{N ^{2/3}}{\sigma_{N}} \,  ( \lambda_1(V_N^G)-\rho_{N}
) \leq x\Bigr)=F_{\rm GUE}(x).}$
\end{itemize}
 \et
\brem \label{rem ExtK} The joint distribution of the $K$ largest
eigenvalues ($K\leq k$ in $(i)$) can easily be
deduced by a straightforward extension of
the arguments of \cite{BaikGBAPeche} in cases $(i)$ and $(iii)$. In case $(i),$ the joint
distribution of the $K$ (correctly rescaled) largest 
eigenvalues $\lambda_i(V_N^G),1 \leq i \leq K, $ converges to the law of the $K$ largest eigenvalues of $H$ (see also \cite{PecheHDR}).
In case $(iii)$, the full conclusion of
Theorem \ref{theo: fluctuwhite} holds true. 
\erem 

Some extensions of Theorem \ref{theo: BaikGBAPeche} have been obtained. First, in \cite{Onatski}, the counterpart of Theorem \ref{theo: BaikGBAPeche} has been established for singular Wishart matrices. Therein, the case where $p <N$ and $\lim_{N \to \infty} \gamma_N= \gamma\in [0,1]$ is investigated. The same phase transition phenomenon is established. 
In \cite{Debashis}, real Wishart ensembles are considered. Unlike complex (singular or not) Wishart ensembles, the joint eigenvalue density is not known in the real setting. Using perturbation theory, it is proved that
when $\pi_1>w_c$ is simple, the largest eigenvalue of real non-white
Wishart matrices $V_N^G$ exhibits Gaussian fluctuations (with a
different variance). Some more recent extensions have also been obtained in \cite{BaiYao} and are recalled below. 

\paragraph{}In view of Theorem \ref{theo: fluctuwhite}, it is natural to investigate the question 
whether Theorem
\ref{theo: BaikGBAPeche} (and its real analogue) would actually hold
true for non Gaussian samples. Our main results, exposed below in Theorem \ref{theo: unicomplex} and Theorem \ref{theo: unireal}, answer this universality question.
Before that, a partial answer has been given at the level of a.s. convergence by \cite{BaikSilverstein}. We
partially state here their result.

\bt \label{theo: baiksilverstein} \cite{BaikSilverstein} Let $V_N$ be a complex or real sample covariance matrix
defined by (\ref{def: VN}) with $\Sigma$ given by (\ref{def: Si}). 
Assume that the entries of $X$ are i.i.d. with $\mathbb{E}X_{11}=0,
\quad \mbE |X_{11}|^2=\sigma^2 $ and  $\mbE |X_{11}|^4 < \infty$.
Let $1 \leq k \leq r$ be an integer. Then,\begin{itemize}
\item[(i)] If $\pi_1=\ldots=\pi_k>w_c$ and $\pi_{k+1}<\pi_1$ then $\lambda _1(V_N), \ldots, \lambda_k(V_N)
$ a.s. converge to $\tau (\pi _1)$.
\item[(ii)] If $\pi_1 \leq w_c$ then $\lambda _1(V_N)$ a.s. converges to
$u_+$.
\end{itemize}
\et
Regarding the fluctuations of the largest eigenvalues now,
\cite{BaiYao} determines their asymptotic distribution in the case
where these eigenvalues are well separated from the bulk. They
consider both the complex and real models in the case where the
entries of $X$ are i.i.d. with finite fourth moment. They prove that $(i)$ in Theorem \ref{theo: BaikGBAPeche} holds true for a wide class of non Gaussian samples $X$ only if the rescaling factor $\sigma (\pi_1)$ given in (\ref{tausigmapi1}) is modified to include the ``excess kurtosis'' of the entries of $X$. The excess kurtosis is given by $\frac{\mbE |X_{11}|^4}{\sigma^4}-2$ (resp. $\frac{\mbE |X_{11}|^4}{\sigma^4}-3$)  in the complex (resp. real) case and is zero for Gaussian distributions.  
One may also indicate that their result is stated in a more general
setting than that considered here: in particular,  the true covariance does not need to be diagonal.
To ease the exposition, we give their result with the added condition (\ref{cond4Moment}) which requires the fourth moment of $X_{11}$ to be as in the Gaussian case. 

\bt \cite{BaiYao} \label{theo: baiyao} Assume that the assumptions of Theorem \ref{theo: baiksilverstein} are satisfied with 
\begin{equation}\label{cond4Moment}
\mbE (X_{11}^4) =(1+\beta') \sigma ^4
\end{equation}
where $\beta' =1$
(resp. $\beta'=2$) in the complex (resp. real) case.
Define $$\xi _i(V_N):= \dfrac{\sqrt N}{\sigma (\pi _1)
}\left ( \lambda_i(V_N^G)-\tau (\pi_1)\right ).$$
If $\pi_1=\ldots=\pi_k>w_c$ and $ \pi_{k+1}<w_c$ then the
$N$-limiting distribution of $(\xi _1(V_N), \ldots, \xi _k(V_N))$ is
the joint distribution of the $k$ eigenvalues, ordered in the decreasing order, of 
the GUE (resp. GOE) $H=(H_{ij})_{i,j=1}^k$ with i.i.d. complex (resp. real) standard
Gaussian entries above the diagonal.\et

The approach of \cite{BaiYao}, following ideas of \cite{Debashis}, is mainly based on the fact (contained
in Theorem \ref{theo: baiksilverstein}) that the largest eigenvalues
split from the bulk. The proof relies on some perturbation
theory ideas which allow to see the rescaled largest eigenvalues $\xi _i(V_N)$ as
the eigenvalues of a $k\times k$ random matrix defined in terms of
the resolvent of an underlying white matrix. The
conclusion then essentially follows from a CLT on random
sesquilinear forms, explaining the assumption on the first four moments of the entries $X_{ij}$.

\paragraph{}The question of the universality of the two other regimes in Theorem \ref{theo:
BaikGBAPeche} remains open. This is the gap we here fill in and is
the main result of this note.
\bt \label{theo: unicomplex} Consider the sequence of complex
sample covariance matrices $(V_N)$ defined by (\ref{def: VN}) where the
entries of $X$ satisfy ${\rm{(H_1)-(H_3)}}$ and $\Sigma$ is given by (\ref{def: Si}). Let $1 \leq k \leq r$ be an integer.\\
When $\pi_1 \geq w_c$, assuming furthermore that
\begin{eqnarray*} \label{4emeMoment}
&{\rm(H_4)}& \mbE (\Re e X_{ij})^4 =\mbE (\Im m X_{ij})^4 = 3 \sigma ^4/4, \quad \forall 1 \leq i \leq N, \forall 1 \leq j \leq p,
\end{eqnarray*}
the conclusions of Theorem \ref{theo: BaikGBAPeche} hold true for $V_N$.\\
When $\pi_1<w_c$, the
conclusion of Theorem \ref{theo: fluctuwhite}
is true for $V_N$. 
\et

In the real setting, we prove the universality in the two
non-critical regimes.

\bt \label{theo: unireal} Consider the sequence of real
sample covariance matrices $(V_N)$ defined by (\ref{def: VN}) where the
entries of $X$ satisfy ${\rm{(H_1)-(H_3)}}$ and $\Sigma$ is given by (\ref{def: Si}). Let $1 \leq k \leq r$ be an integer.\\
When $\pi_1 > w_c$, assuming furthermore that
\begin{eqnarray*} \label{4emeMomentprime}
&{\rm(H_{4}')}& \mbE (X_{ij}^4) =3 \sigma ^4, \quad \forall 1 \leq i \leq N, \forall 1 \leq j \leq p,
\end{eqnarray*}
the conclusion of Theorem \ref{theo: baiyao}
hold true for $V_N$. \\
When $\pi_1<w_c$, the conclusion of Theorem \ref{theo: fluctuwhite}
is true for $V_N$. 
\et

As we will explain, the proof of these theorems is based on a
combinatorial method combined with some results on the corresponding
Gaussian model. In fact, our following combinatorial arguments also
cover the real setting of Theorem \ref{theo: unireal} in the
critical case where $\pi_1=w_c$. Thus, our analysis reduces the
universality problem in this case (under the assumption ${\rm(H_{4}'})$) to the knowledge of the asymptotic fluctuations of the largest eigenvalues of the associated
real Gaussian model. Unfortunately the latter result is not known so far.

\subsection{Core of the proof} {\label{Subsec: Core}}
We here give the main ideas of the proof of Theorems \ref{theo:
unicomplex} and \ref{theo: unireal}. We first mainly concentrate on
the complex setting. At the end of this section, we discuss the main
modifications needed to consider the real case.\\
The proof follows essentially the strategy introduced in \cite{PecheWishart} (see also
\cite{Sos} and \cite{SosWish}) and we refer to this paper for most of
the detail. We also refer the reader to \cite{PecheFeral} where the
authors investigate Deformed Wigner matrices which are classical
Wigner matrices decentered by a particular deterministic matrix. The
Deformed Wigner model can be seen as the additive analogue of the
present model. In particular, it exhibits a similar phase transition
phenomenon regarding the asymptotic behavior of the largest eigenvalues. \cite{PecheFeral} establishes the universality of the fluctuations of the largest eigenvalues for non-necessarily Gaussian Deformed Wigner matrices. The approach developed here is close to that of \cite{PecheFeral} and is mainly based on combinatorial arguments.

\paragraph{}Basically, and for each of the three regimes depending on the
value of $\pi _1$, we
compute the leading term in the asymptotic expansion of expectations
(and also higher moments) of traces of high powers of $V_N$ that is
$\mathbb E ({\rm Tr } V_N^{s_N})$ where ${\rm Tr }$ denotes the
classical (un-normalized) trace. We consider specific exponents
$s_N$ which depend on the scaling of the fluctuations of
the largest eigenvalue(s) when the size $N$ goes to $\infty$. The core
of the proof is to show the universality of moments (of any fixed order) of traces of
powers of $V_N$ in these scales. Let us explain this
more precisely in the particular case of the expectation. 

\indent In the case where $\pi_1 >w_c$, it is expected that the
largest eigenvalue(s) exhibits fluctuations in the scale $N^{-1/2}$
around $\tau (\pi _1)$.
We thus consider an arbitrary sequence of integers 
$(s_N)$ such that $\lim_N s_N/ N^{ 1/2} = c$ for some constant $c > 0$.
We first show that $\mathbb E ({\rm Tr }(V_N/\tau(\pi_1))^{s_N})$ is bounded. Then, we prove that the leading term in the asymptotic expansion of
$\mathbb E ({\rm Tr }V_N^{s_N})$ depends on $\pi _1$, on the
variance $\sigma ^2$ and on the fourth moment of the $X_{ij}$'s only.
Assume now that $({\rm H_4})$ is satisfied i.e. the fourth moment of the $X_{ij}$'s is taken to be
that of the Gaussian distribution with variance $\sigma^2$. Then, up to a negligible error, the 
expectation $\mathbb E ({\rm Tr }V_N^{s_N})$ does not depend asymptotically on the particular law of
the entries and one has that
\begin{equation} \label{ExpUniv}
\mathbb E ({\rm Tr } V_N^{s_N}) = \mathbb E [{\rm Tr } (V_N^G )
^{s_N} ](1 + o(1)).
\end{equation}

In the critical case where $\pi_1 =w_c$, the largest eigenvalue fluctuates now in the scale
$N^{-2/3}$ around the right-edge $u_+$ of the Marchenko-Pastur support. The powers $s_N$ to be considered are such that $\lim_N
s_N/ N^{ 2/3} = c$ for some constant $c> 0$ and we first show that $\mathbb E ({\rm Tr }(V_N/u_+)^{s_N})$ is bounded. We then prove that the leading term in the asymptotic expansion
of $\mathbb E ({\rm Tr }V_N^{s_N})$ depends on $\sigma^2$ and on the fourth moment of the entries of $X$ only. So assuming again that $({\rm H_4})$ holds true, we show that the expectation behaves in the large $N$-limit as in the Gaussian case and that $(\ref{ExpUniv})$ still holds true.

\indent In the sub-critical case $\pi_1 <w_c$, we still consider powers $s_N$
in the order of $N^{ 2/3}$. Here, the fluctuations of the largest
eigenvalues are expected not to depend on $\pi_1$ and to be exactly
as in the white case where $\Sigma =Id$. We get this by showing that $\mathbb E ({\rm Tr} V_N^{s_N}) = \mathbb E [{\rm Tr } M_N  ^{s_N} ](1 + o(1))$.
Thanks to the investigations of \cite{PecheWishart} on the
universality of the fluctuations of the largest eigenvalues of the
white matrices $M_N$, we can deduce that
\begin{equation} \label{ExpUnivBis}
\mathbb E ({\rm Tr } V_N^{s_N}) = \mathbb E [{\rm Tr } (M_N^G )
^{s_N} ](1 + o(1)).
\end{equation}

Actually, we prove universality of all the moments (of fixed order) of
traces of high powers of $V_N$ as in (\ref{ExpUniv}) and (\ref{ExpUnivBis}). Using the
machinery developed in \cite{Sos} (Sections 2 and 5) and
\cite{SosWish} (Section 2), we can then deduce that the limiting
distribution of the largest eigenvalue(s) for spiked population
matrices $V_N$ satisfying $({\rm H_1})$--$({\rm H_3})$ as well as $({\rm H_4})$ if $\pi _1
\geq w_c$, is the same as for complex non-white Wishart matrices $V_N^G$.
When $\pi _1 < w_c$, we more generally get the universality of the
limiting joint distribution of any fixed number of largest eigenvalues. Let us
roughly give the main ideas. On the one hand, the Laplace transform
of the joint distribution of a finite number of the (correctly rescaled) largest eigenvalues of $V_N$ can be conveniently expressed in terms of joint moments of traces of the matrix $V_N$ taken at
suitable powers $s_N$ (those of the
previous discussion). On the other hand, the asymptotic distribution
of the rescaled largest eigenvalues (and also the corresponding
Laplace transform) is well-known in the complex Gaussian setting.
One can then deduce from universality of moments of traces that the
asymptotic joint distribution of the largest eigenvalues for any
model $V_N$ considered here is the same as for the corresponding
Gaussian case. The detail of the derivation of such a result from
formulas (\ref{ExpUniv}) and (\ref{ExpUnivBis}), including the required asymptotics of
correlation functions for the complex non-white Wishart matrix $V_N^G$, can
be found in \cite{PecheFeral}, \cite{BaikGBAPeche}, \cite{Sos} and
\cite{SosWish}.

\paragraph{} In the real setting, our combinatorial reasoning also yields the universality of moments of
traces of high powers of $V_N$. This could be used in principle to prove universality of the fluctuations of the largest eigenvalues, provided the full counterpart of Theorem \ref{theo: BaikGBAPeche} in the real Gaussian case was fully established. 
This is true in the case where the largest eigenvalue of the true covariance is simple and satisfies $\pi _1
>w_c$ (cf. \cite{Debashis}). The non Gaussian case is actually also covered by Theorem \ref{theo: baiyao}. We can come to the same conclusion (assuming $({\rm H'}_4)$) using our approach.
In the sub-critical case where $\pi _1 <w_c$, we are also able to conclude thanks to
(\ref{ExpUnivBis}) (and its analogue for higher moments) which
proves that the fluctuations of eigenvalues of $V_N$ are similar to
those, well-known, of the real Wishart matrix $M_N^G$. In the critical case
$\pi _1 =w_c$, we cannot conclude.

\paragraph{}Our paper is organized as follows. In Section \ref{sec:
  combi}, we introduce the major combinatorial tools needed to compute
moments of traces of high powers of $V_N$. We first recall the
specific terminology and the main arguments (Section \ref{sec: white})
developed by \cite{PecheWishart} for the investigations of the white
case. We then present (Section \ref{sec: counting}) the main ideas of
the strategy we will use to deal with the non-white case when $r=1$. In Section \ref{sec: expect}, we establish the universality of the asymptotic expectation of traces of high powers of $V_N$. We next consider higher moments in Section \ref{sec: higher}. 
In Section \ref{Sec: r>1}, we discuss the main modifications needed to deal with the case where $r>1$.

\paragraph{}{\it Acknowledgments.} A part of this work was done while the first
author prepared her PhD Thesis at the Institut de Math\'ematiques de Toulouse and she acknowledges useful conversations with
her advisor M. Ledoux. Some results were also obtained last year during a postdoctoral fellow
at the Hausdorff Research Institute for Mathematics of Bonn.

\section{Combinatorial tools} {\label{sec: combi}}
In this section, we define the major combinatorial tools needed to investigate the asymptotics of moments of traces of large powers of $V_N.$ We here extend some of the tools used in \cite{PecheWishart} where the white case ($\Sigma=Id$) is investigated. The reader is referred to Sections 2 and 3 of the above cited article for a detailed explanation of the following combinatorial approach. We here choose to explain our strategy in the case where $r=1$ and thus consider the covariance matrix 
$$\Sigma=\text{diag}(\pi_1, 1, 1, \ldots, 1).$$
Modifications to handle more complex cases ($r>1$) are indicated in
Section \ref{Sec: r>1}.\\
Thoughout the paper, we denote by $C,C', C_i, i=1,2, \ldots$ some positive
constants independent of $N$ and whose value may vary from line to line.

\subsection{Paths and $1$-edges}
Let $(s_N)$ be a sequence of integers that may grow to infinity. Developing the trace, one obtains that
\begin{eqnarray}
&&\mbE \big [ {\rm{Tr}} \, V_N^{s_N} \big ] \label{tracemn}\\
&&=\frac{1}{p^{s_N}}\sum_{i_0,i_1, \ldots, i_{s_N-1}}
\pi_1^{\sum_{q=0}^{s_N-1}\delta_{1i_q}}\,\mbE \left
(\prod_{q=0}^{s_N-1} (XX^*)_{i_qi_{q+1}}\right)\crcr &&=
\frac{1}{p^{s_N}}\sum_{i_0,i_1, \ldots, i_{s_N-1}} \pi_1^{\# \{q: \,
i_q=1\} }\sum_{j_1, \ldots,j_{s_N}}\,\mbE \left (\prod_{q=0}^{s_N-1}
X_{i_qj_{q+1}}\overline{X_{i_{q+1}j_{q+1}}}\right)
\label{tracemndev}\\
&& \text{ where $i_q \in \{1, 2, \ldots, N\}$ \quad and \quad
$j_{q} \in \{1, 2, \ldots, N, \ldots, p\}$} \label{Rule}
\end{eqnarray}
and where we use the convention that $i_{s_N}=i_0$.

\paragraph{} To each term $\prod_{q=0}^{s_N-1}
X_{i_qj_{q+1}}\overline{X_{i_{q+1}j_{q+1}}}$ in
$(\ref{tracemndev})$, we associate three combinatorial objects
needed in the following. First, we define the "edge path" $\mathcal
P$ formed with oriented edges (read from bottom to
top) by \be \mathcal{P}=\begin{pmatrix}j_1\\ i_0 \end{pmatrix}
\begin{pmatrix}j_1\\ i_1 \end{pmatrix}
\begin{pmatrix}j_2\\ i_1 \end{pmatrix}\cdots  \begin{pmatrix}j_{s_N}\\ i_{s_N-1}\end{pmatrix}
\begin{pmatrix}j_{s_N}\\  i_0 \end{pmatrix}.\label{edgepath}
\ee Due to the symmetry assumption ${\rm(H_3)}$, only paths for
which any oriented edge occurs an even number of times give a non
zero contribution to the expectation. From now on, we only consider
such even paths.\\
To an arbitrary even edge path $\mathcal P $, we associate a
so-called Dyck path (Dyck paths have a long history in Random Matrix
Theory, see \cite{Wigner}, \cite{Bai} for instance), that is a
trajectory $x= \{ x(t), t\in[0, 2s_N]\}$ on the positive
half-lattice such that:$$ x(0)=x(2s_N)=0; \quad  \forall t \in [0,
2s_N], \, x(t)\geq 0 ~~ \text{and} ~~ x(t)-x(t-1)=\pm 1.$$ To
define the Dyck path $x$ associated to $\mathcal P $, we read the
oriented edges of $\mathcal P $ in the order of appearance and draw
an up (resp. down) step $(1,+1)$ (resp. $(1,-1)$) if the current
edge is read for an odd (resp. even) number of times. Last, we also
associate to $\mathcal P $ a "usual" path denoted by $P$ :
we mark on the underlying Dyck path $x$ the successive vertices met
in $\mathcal P $ and then set $P:=i_0 \, j_1 \, i_1 \, j_2 \ldots i_0.$

\paragraph{} The strategy in the rest of the paper can roughly be summarized as follows. Given a trajectory $x$, we shall estimate the number of edge paths that can be associated to $x$ and then we shall estimate their contribution to the
expectation (\ref{tracemn}). On the one hand, due to the constraint
(\ref{Rule}) on the choice of the vertices, we shall refine the
enumeration of Dyck paths according to the number of odd up steps. The way to handle such a specificity has been developed in detail in \cite{PecheWishart}. We recall some points of the analysis made by \cite{PecheWishart} in the next subsection. On the other hand, when estimating the
contribution of such edge paths, we also have to take into account the occurrences of the vertex $1$
on the bottom line since each occurrence yields an additional weight $\pi _1$ (recall $(\ref{tracemndev})$). To this aim, we introduce the notion of
$1$-edge.

\bdefi A $1$-edge is an oriented edge with $1$ on the bottom line
i.e. an edge $\begin{pmatrix}h\\ 1
\end{pmatrix}$ where $h \in \{1, \ldots, p\}.$ \edefi \brem An edge
$\begin{pmatrix}1\\ h \end{pmatrix}$ is not a $1$-edge in our
denomination if $h\in \{2, \ldots, N\}.$ \erem

Thus, we shall be able to refine the analysis made in
\cite{PecheWishart} to
estimate the contribution of edge paths with 1-edges. \\
The rest of this section is organized as follows. In Subsection \ref{sec: white}, we recall the main definitions and
results of \cite{PecheWishart} that we will use throughout this
paper. Note
that the investigations of \cite{PecheWishart} readily gives the contribution to the expectation (\ref{tracemn}) of edge paths without 1-edges (see Proposition \ref{Propcontriwithout1edge}). In Subsection \ref{sec: counting}, we explain the main
ideas of the strategy we will use to deal with paths with 1-edges
and compute the corresponding contribution to the expectation
(\ref{tracemn}).

\subsection{The white case $\Sigma=Id$ and paths with no $1$-edges\label{sec: white}}
The aim of this section is to recall the main definitions and
results derived from \cite{PecheWishart} that we will use throughout
this paper. These results also allow us to estimate at the end of
this subsection the contribution to (\ref{tracemn}) of paths without
$1$-edges. We assume some familiarity of the reader with the combinatorial
machinery developed in refs. \cite{Si-So1}, \cite{Si-So2},
\cite{Sos} and \cite{PecheWishart}.

\begin{NotRem} Our notations differ from those used in \cite{PecheWishart} :
 the (white) model considered by \cite{PecheWishart} corresponds here to $\frac{\sigma
^{-2}}{N} X X^*$ that is $\sigma ^{-2} \gamma_{N} M_N$.
\end{NotRem}

\paragraph{} To handle the case where $\Sigma=Id$ (that is the computations of $\mbE \tTr M_N^{s_N}$), \cite{PecheWishart} first
enumerates the associated Dyck paths according to the number of odd
up steps. Thus, we let $\chi_{s_N,k}$ be the set of Dyck paths of
length $2s_N$ with
$k$ odd up steps and $\chi_{s_N}=\cup_{i=1}^{s_N} \chi_{s_N,k}$ be
the set of Dyck paths of length $2s_N$.
\bdefi \label{Prop:
Narayana}\cite{Chen} Let $\mathbf{N} (s_N, k)$ be the $k$th Narayana number defined by \be
\label{Narayananumber}\mathbf{N}(s_N,k)=\frac{1}{s_N}\binom{s_N}{k}
\binom{s_N}{k-1}.\ee Then $\mathbf{N}(s_N,k)=\sharp
\mathcal{X}_{s_N,k}.$ \edefi \noindent The reader is referred to
\cite{PecheWishart} and references therein for further detail about
Narayana numbers.

\paragraph{}Given a Dyck path $x \in \mathcal{X}_{s_N,k}$ we shall estimate the number of edge paths associated to it. First, one
needs to assign a vertex from $\{1, \ldots, N\}$ (resp. $\{1,\ldots,
p\}$) to each even (resp. odd) moment of time along the Dyck path
$x$. For this, we need the following definition.

\bdefi An instant $t\in [1,2s_N]$ is said to be marked (in $x$ or in
$P$) if it corresponds to the right endpoint of an up edge. \edefi

Roughly speaking, marked instants correspond to the moments of time
(apart from $t=0$) where one can discover in $P$ a vertex never
encountered before. Thus, in order to estimate the number of paths
that can be associated to a given trajectory $x$, we first
choose the vertices occurring at the marked instants, which we call
marked vertices, and at the origin of the path (which is non marked
by definition). \\
In order to choose the vertices occurring at the marked instants, we
refine our classification separating the cases where they
are on the bottom or top line in $\mathcal P $, that is the cases
where they are marked at even or odd instants in $P$, as follows. For any integer $0\leq i \leq s_N$, we define two classes
of marked vertices: $\mathcal{N}_i=\{\text{ vertices occuring $i$ times at an even
marked instant }\}$ and $\mathcal{T}_i=\{\text{ vertices occuring
$i$ times at an odd marked instant }\}$
and we set $n_i=\sharp \mathcal{N}_i$ and $p_i=\sharp \mathcal{T}_i$. Then (cf. (\ref{Rule})), vertices encountered along a path $P$ at the odd (resp. even) instants split into the disjoint classes $\mathcal {T}_i$ (resp. $\mathcal {N}_i$).
Observe that $n_i=0$, $\forall i>s_N-k$ with $\sum _{i }  n_i=N$
and $\sum _{i } i n_i=s_N-k$. Similarly, one has that $p_i=0$, $\forall
i>k$ with $\sum _{i }  p_i=p$ and $\sum _{i } i p_i=k$.  A path $P$ can then be characterized by its associated trajectory $x\in \chi_{s_N,k}$ for some integer $1\leq k \leq s_N,$ and its type:
$$(n_0, n_1, \ldots, n_{s_N-k})(p_1, \ldots, p_k):=(\tilde n , \tilde p).$$

\bdefi A vertex $v \in \mathcal{N}_i $ (resp. $v \in \mathcal{T}_i$)
is said to be of type $i$ on the
bottom (resp. top) line.\\
Any vertex $v \in \cup _{i \geq 2} \mathcal{N}_i $ (resp. $v \in
\cup _{i \geq 2} \mathcal{T}_i$) is said to be a vertex of
self-intersection on the
bottom (resp. top) line.\edefi

In the following, $M_1=\sum _{i \geq 2} (i-1) n_i$ (resp. $M_2=\sum _{i \geq 2} (i-1)
p_i$) denotes the number of vertices of self-intersection on the
bottom (resp. top) line.

\paragraph{}It is an easy fact that, given the type $(\tilde n , \tilde p)$, the
number of ways to assign vertices at the marked instants and choose
the origin is at most 
\begin{eqnarray}\label{boundchoicesvertices}
N \frac{N!}{\prod_{i=0}^{s_N-k} n_i!} \frac{p!}{\prod_{i=0}^{k}
p_i!} \frac{(s_N-k)!}{\prod_{i \geq 2 } (i!)^{n_i}}
\frac{k!}{\prod_{i \geq 2 } (i!)^{p_i}}.
\end{eqnarray}
Once marked vertices are chosen, there remains to count the number of ways to fill in the blanks of the path i.e. assign
vertices at the unmarked instants and evaluate the corresponding expectation of each ``filled path''. Due to self-intersections, there may be many ways to fill in the blanks of the path as well as edges seen many times. Actually the bound (\ref{boundchoicesvertices}) would be enough for the following as long as $s_N =o(\sqrt{N}).$ It needs to be refined for higher scales $s_N.$ In particular and as explained in the beginning of Section 3.2 in \cite{PecheWishart}, one must pay attention to vertices of type 2. In (\ref{boundchoicesvertices}), we used the rough estimate that if $t\in \{1,\ldots, k\}$ (resp. $t\in \{1,\ldots, s_N-k\}$) is an odd (resp. even) marked instant where the second occurrence of a vertex of type $2$ is repeated, there are at most $t-1$ possible choices for the vertex to be repeated. 
This rough estimate needs to be refined when considering vertices $v$ of type 2 belonging to edges seen more than twice and vertices $v$ of type 2 for which there are multiple ways to close an edge with $v$ as its left endpoint at an unmarked instant.\\
Let us first consider the latter class of vertices of type $2$. Note that for such a vertex $v$, there are at most three possible ways to close an
edge with $v$ as left endpoint at an unmarked instant. To investigate this class, we need a few definitions from
\cite{PecheWishart}.

\bdefi A vertex $v$ of type 2 is said to be non-MP-closed if it is an odd (resp. even) marked instant and if there is an ambiguity for closing an edge at an unmarked instant starting from this vertex on the top (resp. bottom) line.
\edefi
\noindent Let $t$ be a given marked instant. Assume that the marked vertices before $t$ have been chosen and that, at the instant $t$, there is a non-MP-closed vertex. Then, by definition of $x$ and of non-MP-closed vertices, there are at most $x(t)$ possible choices for this vertex. This can be checked as in \cite{Si-So2}, p. 122.
\par Let us turn to vertices of type $2$ which belong to an edge that is read
four times or more in the path. To consider such vertices, we need to introduce other characteristics of
the path. Let $\nu _N:=\nu_N( P)$ be the maximal number of vertices that can be visited at
marked instants from a given vertex of the path $P.$ Let also $T_N:=T_N(P)$ be the maximal type of a
vertex in $P.$ Then, if at the instant $t$, one reads for the second time an
oriented up edge $e$, there are at most $2(\nu_N+T_N)$ choices for the vertex occurring
at the instant $t$. Indeed, one shall look among the oriented edges
already encountered in the path and for which one endpoint is the
vertex occurring at time $t-1$ (see the Appendix in \cite{Si-So2} and
Section 5.1.2 in \cite{PecheFeral} e.g.).

\paragraph{} Furthermore, the machinery developed by \cite{Si-So1},
\cite{Si-So2}, \cite{Sos} and \cite{PecheWishart} shows that once
the marked vertices are assigned along a trajectory $x$ and once the
origin is chosen, the number of filled paths, weighted by their expectation, which are associated to $x$ (and of type $(\tilde n , \tilde p)$) is bounded by
\begin{equation}\label{OmegaNmbEN}
\frac{2 \sigma ^{2s_N}}{p^{s_N}}\prod_{l= 3}^{s_N-k}\left (
C_1l\right)^{ln_l}\prod_{m= 3}^{k}\left
(C_1m\right)^{mp_m}3^{r_1+r_2}C_2^{q_1} C_3^{q_2},\end{equation} where the extra factor $2$ comes from the negligible case where the
origin $i_0$ is marked; the $C_i$'s are positive constants independent of $k, p, N$ and
$s_N$; $r_i, i=1, 2$ (resp. $q_i, i=1,2$) count the number of
vertices of type 2 on the bottom/top line which are non-MP-closed
(resp. belong to an edge seen more than twice).

\paragraph{}Combining the above with some ideas previously developed in the above cited papers, it is shown in Section 3 in \cite{PecheWishart} that 
the contribution to $\mbE \tTr M_N^{s_N}$ from paths with
$k$ odd marked instants and of type $(\tilde n , \tilde p)$ is
bounded by
\begin{eqnarray}\label{gencontrk}
&&  C\mathbf{N}(s_N, k)  N  \gamma_N^{k-s_N}  \sigma ^{2s_N}
\sum_{r_1, r_2,q_1,q_2, n_i,p_i,i\geq 2}
e^{\{-\frac{(s_N-k-M_1)^2}{2N}-\frac{(k-M_2)^2}{2p}\}}\mathbb{E}_{k}\Big [\crcr &&
\frac{\Bigl ( 3(s_N-k)\max
\:x(t)\Bigr)^{r_1}}{r_1!}\frac{\Bigl (C_1(s_N-k)(\nu_N +T_N)
\Bigr)^{q_1}}{q_1!}\prod_{i\geq 3}\frac{\left (
\frac{{C}^i(s_N-k)^i}{N^{i-1}} \right)^{n_i}}{n_i!} \crcr && ~~~~~~
\frac{\left(\frac{(s_N-k)^2}{2}\right)^{n_2-r_1-q_1}}{(n_2-r_1-q_1)!} \: \frac{\Bigl ( 3k\max
\:x(t) \Bigr)^{r_2}}{r_2!}\crcr && ~~~~~~
\frac{\left( \frac{k^2}{2}\right)^{p_2-r_2-q_2}}{(p_2-r_2-q_2)!}\:
\frac{\Bigl (C_2k (\nu_N +T_N)\Bigr)^{q_2}}{q_2!}\: \prod_{i\geq
3}\frac{1}{p_i!} \left (\frac{{C}^ik^i}{p^{i-1}} \right)^{p_i}\Big ],
\end{eqnarray}
where\\
- $C$, $C_1$ and $C_2$ are positive constants independent of $k, \, p, \, N$ and
$s_N$;\\
- $\max x(t)$ is the maximal level reached by a trajectory $x$; \\
- $\mbE_{k}$ is the expectation with respect to the uniform distribution on $\mathcal{X}_{s_N,k}$.\\
Moreover, \cite{PecheWishart} (Section 3) establishes important
estimates on the previous quantities. Proposition 3.1 in \cite{PecheWishart} proves that typical paths (that is paths which contribute
in a non negligible way to $\mbE \tTr M_N^{s_N}$) satisfy the following
constraints:
\begin{itemize}
\item [a)] the number $k$ of odd marked instants lies
in the interval $[\alpha' s_N, \alpha s_N]$ for any $\alpha',
\alpha$ such that $0<\alpha'<\frac{\sqrt \gamma}{1+\sqrt
\gamma}<\alpha<1$; 
\item[b)] $\nu_N+T_N<<\sqrt{s_N}$;
\item[c)] there exists a constant $c>0$, independent
of $k, \, p, \, N$ and $s_N$, such that $M_1+M_2 \leq c \sqrt
{s_N}$.            
\end{itemize}
Besides, \cite{PecheWishart} also proves that $\max x(t)\sim
\sqrt{s_N}$ in typical paths. More precisely, it is shown in Lemma 3.1 in
\cite{PecheWishart} that
\begin{equation}\label{estimexpMaxPe}
\forall a >0, \, \exists C(a)<\infty, \, \max_{\alpha' s_N\leq k\leq \alpha s_N}\mbE_{k} \left [ \exp{\{a \max
x(t)/\sqrt{s_N}\}}\right]\leq C(a).
\end{equation}

\paragraph{ } All these results combined with (\ref{gencontrk}) lead to one of
the main results of Section 3 in \cite{PecheWishart}.

\begin{prop}If $\Sigma=Id$ and $s_N=O(N^{2/3})$, the paths with $k$
odd marked instants contributing to $\mbE \tTr M_N^{s_N}$
in a non-negligible way have edges read only twice, a non marked origin and no vertex of type strictly greater than 3.
Furthermore, there exists a constant $C>0$ such that their contribution is at most
\begin{equation}\label{contriwhiteSandrine}
 \mathbf{N}(s_N, k) \, N \, \gamma_N^{k-s_N} \, \sigma ^{2s_N} \, \mathbb{E}_{k}\Big [\exp \Big ( 6
\frac{s_N \max _t x(t) }{N} \Big) \Big ] \exp \Big ( C
\frac{s_N^3}{N^2} \Big ).
\end{equation}
\end{prop}

\paragraph{ }From these computations, one readily deduces that the contribution to the expectation (\ref{tracemn}) from paths
without 1-edges is characterized as follows.

\begin{prop} \label{Propcontriwithout1edge}
The typical contribution to the expectation (\ref{tracemn}) from paths with no 1-edges is at most of the order of
\begin{eqnarray*}
\sum_{\alpha' s_N\leq k\leq \alpha s_N}\mathbf{N}(s_N, k) \, N \,
\frac{\gamma_N^{k}}{\gamma _N ^{s_N}} \, \sigma ^{2s_N} \, \mathbb{E}_{k}\Big [\exp \Big ( 6
\frac{s_N \max _t x(t) }{N} \Big) \Big ] e^{ C \frac{s_N^3}{N^2} } &\leq & C' u_+^{s_N},\end{eqnarray*} 
where $C$ and $C'$ are positive constant independent of $N$.
Typical
paths amongst those without 1-edges have edges read only twice, a
non marked origin and vertices of type at most 3.
\end{prop}

\paragraph{Proof of Proposition \ref{Propcontriwithout1edge}:} Here the vertices
$i_j$'s must be chosen from the set $\{2, \ldots, N \}$ instead of $\{1, \ldots, N \}$, since the
vertex $1$ is assumed not to occur on the bottom line. Formula
(\ref{gencontrk}) must be simply multiplied by a factor $(1-N^{-1})^{s_N-k+1}.$
This has no impact on the final result for $N$ large enough.
Proposition \ref{Propcontriwithout1edge} follows by summation on
the typical $k$'s and the fact that $\sum_{k=1}^{s_N}\mathbf{N}(s_N,
k) \, N \, \gamma_N^{k-s_N} =O \left ( (u_+/\sigma^2 )^{s_N}\right)$ (see Remark 2.4 in \cite{PecheWishart}). $\square$

\begin{NotRem} From now on, we simplify the notations and use $P$ to denote a
usual path as well as its associated edge path $\mathcal P$.
\end{NotRem}

We shall now be able to refine the counting procedure of
\cite{PecheWishart} to estimate the contribution to the expectation (\ref{tracemn}) from edge paths
(\ref{edgepath}) with 1-edges. The problem of evaluating directly
the number of $1$-edges occurring in a path turns out to be
difficult. Thus, our strategy will be indirect. Instead of directly
evaluating the contribution of paths $P$ as well as the number of
its $1$-edges, we first evaluate the contribution of paths with a
prescribed number of $1$-edges. This is the aim of the following
subsection.

\subsection{Counting the number of $1$-edges} \label{sec: counting}

In this subsection, we define a procedure called {\it gluing
procedure} which allows to enumerate the paths $P$ according to the
number of their $1$-edges. One can first notice that the
contribution of a path $P$ with $1$-edges to the expectation (\ref{tracemn}) is a
weighted term related to its contribution to the expectation $\mbE[
\text{Tr}M_N^{s_N}]$, where $M_N=\frac{1}{p}XX^*$ is the associated white matrix. One simply assigns
a weight $\pi_1$ to each occurrence of the vertex $1$ on the bottom
line of $P$. Consider for a while a path $P$ of length $2s_N$ having $s \geq 1$ pairs of 1-edges and with $1$ for origin. The basic idea
is that the vertex $1$ necessarily occurs on the bottom line of $P$
at the instants where the trajectory $x$ of $P$ hits the level $0$. If one
furthermore assumes that $x$ hits exactly $s$
times the level 0, then all the occurrences of $1$ on the bottom
line of $P$ correspond to the returns to $0$ of $x$. Thus the enumeration
of $1$-edges transfers to statistics on the number of returns to $0$
for Dyck paths. This observation is the basic idea of the gluing
procedure.

Throughout the paper, we denote by $2s$ the number of 1-edges of a path $P$.
Starting from a general path $P$ of length $2s_N$, the gluing procedure associates a new path $P'$ with origin $1$ as we now explain.

\subsubsection{Subpaths starting and ending with a $\mathbf 1$-edge}
We denote by $\begin{pmatrix} h_i\\1 \end{pmatrix}$ $\begin{pmatrix}
g_i\\1 \end{pmatrix}$, $i=1, \ldots,s$ the pairs (not necessarily
distinct) of successive $1$-edges occuring in $P$. One can then write
$P$ as
\begin{eqnarray*}
&&P=\underbrace {\begin{pmatrix}j_1\\ i_0 \end{pmatrix} \begin{pmatrix}j_1\\ i_1 \end{pmatrix}\cdots \begin{pmatrix}h_1\\ 1 \end{pmatrix}}\underbrace{\begin{pmatrix}g_1\\ 1 \end{pmatrix}\cdots
\begin{pmatrix}h_2\\ 1 \end{pmatrix}}\underbrace{\begin{pmatrix}g_2\\ 1 \end{pmatrix}\cdots}\cdots \underbrace{\cdots \begin{pmatrix}h_s\\ 1 \end{pmatrix}}  \cr && ~~~~~~~~~ \underbrace{\begin{pmatrix}g_s\\ 1 \end{pmatrix} \cdots\begin{pmatrix}j_{s_N}\\ i_{s_N-1}\end{pmatrix}
\begin{pmatrix}j_{s_N}\\  i_0 \end{pmatrix}}.
\end{eqnarray*}
Using these $1$-edges, $P$ splits into $s$ subpaths $P_i,
i=1, \ldots, s,$ defined as follows. For $i=2, \ldots, s,$ we call $P_i$
the subpath starting at $\begin{pmatrix} g_{i-1}\\1
\end{pmatrix}$ and ending at $\begin{pmatrix} h_i\\1 \end{pmatrix}.$
Let then $P_1$ be the subpath $\begin{pmatrix}g_s\\ 1 \end{pmatrix}
\cdots\begin{pmatrix}j_{s_N}\\ i_{s_N-1}\end{pmatrix}
\begin{pmatrix}j_{s_N}\\  i_0 \end{pmatrix}.\begin{pmatrix}j_1\\ i_0 \end{pmatrix} \begin{pmatrix}j_1\\ i_1 \end{pmatrix}\cdots \begin{pmatrix}h_1\\ 1 \end{pmatrix}.$

\brem In the particular case where $i_0=1$, the edge path reads as
\begin{eqnarray*}
&&P=\underbrace {\begin{pmatrix}j_1\\ 1 \end{pmatrix} \begin{pmatrix}j_1\\ i_1 \end{pmatrix}\cdots \begin{pmatrix}h_1\\ 1 \end{pmatrix}}\underbrace{\begin{pmatrix}g_1\\ 1 \end{pmatrix}\cdots
\begin{pmatrix}h_2\\ 1 \end{pmatrix}}\underbrace{\begin{pmatrix}g_2\\ 1 \end{pmatrix}\cdots}\cdots \underbrace{\cdots \begin{pmatrix}h_s\\ 1 \end{pmatrix}}.
\end{eqnarray*}
The sole difference here is that $g_s=j_1$ and $P_1$ is the subpath beginning $P$.
\erem

\paragraph{} Let $t_1, \ldots, t_s$ denote the instants at which the
successive $s$ pairs of $1$-edges occur in $P.$
The length $l_i$ of each subpath is determined by the
instants $t_i$ since $l_i=t_{i+1}-t_i$ (note that these lengths
are necessarily even).
One can also note that the $1$-edges occurring in the path are
necessarily even. Thus, for any $1\leq i\leq s$, there exists $j$
such that $h_i=h_j$ or $h_i=g_j.$ This observation is crucial for
the sequel. As we will explain in the following
subsection, this allows us to re-order the paths $P_i$ and erase
some of the $1$-edges.

\subsubsection{Gluing and reordering the paths $P_i$\label{subsec: glu}}
Given the set of $1$-edges, we now define a graph $G$ on the set $\mathcal L =\{g_{i-1}, h_{i}, i=1, \ldots s\}$ ($\# \mathcal L \leq s$) of the
vertices occurring in $1$-edges, using the convention that
$g_{0}=g_s$. We draw an edge between $g_{i-1}$ and $h_{i}$ for $i=1,
\ldots, s$. Note that multiple connections are allowed. We denote by $l,1 \leq  l \leq s,$ the number of connected components of
$G$. We then group all together the subpaths associated to vertices
of the same connected component of $G$. This leads to $l$ subsets
which we call {\it clusters}. Clusters are ordered in the order they are encountered in $P$ and we denote them by $\mathcal S_j,\, 1 \leq j\leq l$.

\paragraph{}These clusters will now be used to build a new path $P'$ from $P$ as follows. We first define a way to glue the subpaths belonging to the same cluster. For any $1 \leq j \leq l$, we will denote by $P_j^g$ the final path obtained by the gluing of the subpaths from the cluster $\mathcal S _j$.
\begin{itemize}
\item Assume first that $\sharp \mathcal L=s$ so that
each $1$-edge occurs exactly twice in $P$. Consider the first
cluster which, by definition, begins with the subpath $P_1$. We
first read $P_1$ until meeting the edge $\begin{pmatrix} h_1\\1
\end{pmatrix}.$ If $g_s=h_1$ (then $\mathcal S _1$ contains only $P_1$), then the process stops and the path $P_1^g$ is equal to
$P_1.$ Otherwise, there exists $j_o\geq 2$ such that $P_{j_o}$ has
the edge $\begin{pmatrix} h_1\\1 \end{pmatrix}$ as endpoint. In the
case where $\begin{pmatrix} h_1\\1 \end{pmatrix}$ is the left
endpoint of $P_{j_o}$, we concatenate $P_1$ and $P_{j_o}$ and then
erase the two occurrences of the $1$-edge $\begin{pmatrix} h_1\\1
\end{pmatrix}.$ In the case where the edge $\begin{pmatrix} h_1\\1
\end{pmatrix}$ is the right endpoint of $P_{j_o}$, we read
$P_{j_o}$ in the reverse order and apply a similar procedure. This
``gluing'' defines a new subpath which we denote by $P_1\vee
P_{j_o}$. We then restart the procedure with $P_1$ replaced with
$P_1\vee P_{j_o}$ until all the subpaths belonging to the first cluster
are glued leading to the final subpath $P_1^g$. We then proceed in
the same way with other clusters.
\item If $\sharp \mathcal L<s$ then some clusters have $1$-edges that occur four times or more in $P$. We can find a way to read all the edges of such a cluster without ``raising the pen''. This follows from the fact that the vertices of $G$ are all of even valency. We then choose one way to do so and glue the paths of these clusters
accordingly.  For clusters having $1$-edges that occur only twice, we apply the previous gluing method.
\end{itemize}
We end up with
$l$ paths $P_j^g, j=1, \ldots, l,$ which begin and end with a
$1$-edge. By definition of the clusters, these $1$-edges form
$l$ pairwise distinct pairs of oriented edges. We then call $P'$ the
path obtained by the concatenation of the $P_j^g$ 's that is $P'= P_1^g \cup \ldots \cup P_l^g$. The length of $P'$
is $2(s_N-(s-l))$ and its origin is $1$, which is a non marked vertex on the bottom line. We call $x'$ its trajectory.

\paragraph{} The basic idea of the gluing procedure defined above can be
roughly explained as follows. A path $P$ (or $P'$) is said to be
typical if it contributes in a non-negligible way to the expectation
(\ref{tracemn}). 
We first identify the typical paths $P'$. The simplest of these typical paths 
are such that the number of occurrences of $1$-edges is
determined by the number of returns to $0$ of their associated
trajectory $x'$. Then, given a typical path $P'$,
one has to estimate the number of paths $P$ that can be associated
to it as well as their expectation. When considering the
expectation, we shall take into account the added weight due to the erased 1-edges. This problem will be considered in the following
section. We here establish the needed estimate for the number of
preimages $P$ of a path $P'$.

\subsubsection{Number of preimages of a glued path $P'$}
The simplest case is when $l=s$ since all the preimages $P$ coincide
with $P'$ up to the translation of the origin. Then if the first
return to $0$ of the trajectory $x'$ associated to $P'$ holds at the instant $T=2s_1$, there are exactly (resp. at most) $s_1$ preimages of
$P'$ if $x'$ returns $m=s$ (resp. $m <s$) times to the level 0. In the
other case where $l<s$, the following estimate holds true.

\bl \label{Lemm: nbrpreimagesP'} Assume that the first return to
$0$ of $x'$ holds at time $T=2s_1$.
Then the number of preimages $P$ of the path $P'$ does not exceed
\begin{equation} \label{EstimNbrpreimagesP'}
s_1 \, \binom{s}{l} \, \left ( 2s_N\right )^{s-l}. \end{equation} \el

\paragraph{Proof of Lemma \ref{Lemm: nbrpreimagesP'}:}In order to reconstruct the initial subpaths $P_i, i=1, \ldots , s$ from $P'$, we first need to choose the $s-l$ instants where we have
erased $1$-edges. The set $ \mathcal T$ of these instants combined with
the $l$ occurrences of pairs of $1$-edges in $P'$ (which determine
the paths $P_j^g$) define $s$ subpaths $P_i'$ which are the subpaths $P_i$
possibly read in the reverse direction. Then one has to define the
order in which the subpaths $P_i$ are read. The sole constraint on
this order bears on the path starting each cluster, as we now
explain. Consider for instance a cluster, say $\mathcal
S_j$ in $P$ and its corresponding counterpart $P_j^g$ in $P'$.
Call $t$ the first instant of $\mathcal T$ chosen in $ P_j^g$. Then
the subpath $P'_i$ starting $ P_j^g$ and ending at $t$ is the first
subpath (with the same direction) of the cluster
$\mathcal S_j$
met in $P$. Last and in order to define completely the path $P$, one also has to choose the origin of $P$.
Thanks to the above, we can now show that the number of preimages of
a given path $P'$ does not exceed: \be \label{firstnbrpreP'} s_1
\binom{s_N-1}{s-l} \frac{s!}{l!} 2^{s-l}.\ee It is clear that the previous
binomial coefficient comes from the choice of the $s-l$ instants of
$\mathcal T$ (noticing that these instants are necessarily odd). To explain
the remaining terms in (\ref{firstnbrpreP'}), we denote by $x_j$ the
number of subpaths $P_i$ in each cluster $\mathcal S_j, 1 \leq j
\leq l.$ Then, let us consider the first cluster $\mathcal S _1$. As
clusters are interlaced in $P$, we also have to choose the places
where we read the $x_1-1$ paths of $\mathcal S _1$ (different from
$P_1$) and choose the order in which we read them. There are
$\binom{s-1}{x_1-1} (x_1-1)!$ such choices. Furthermore one can also
choose the direction in which one reads each of the $P'_i$ (to
obtain $P_i$) not beginning ${P}_1^g.$ There are two choices
for this direction. Having done so, the first empty
``slot'' corresponds necessarily to the time where we read the first
path of the second cluster. We use the same procedure to define the
order and the direction in which the remaining subpaths of the
second (and subsequent) clusters are read. Thus the number of ways
to determine and reorder the subpaths $P_i$ is at most
$$ \binom{s_N-1}{s-l} \prod_{j=1}^l  \binom{s-1-\sum_{j'=0}^{j-1}x_{j'}}{x_j-1} \, (x_j-1)! 2^{s-l} \leq \binom{s_N-1}{s-l} \, \frac{s!}{l!} 2^{s-l},$$
where we took the
convention that $x_0=0$. Last, we shall add a term $s_1$ coming from the
determination of the origin $i_0$ of the initial path $P$ (which amounts to choosing a vertex occurring on the bottom line of $P_1$). This yields (\ref{firstnbrpreP'}) and it is then easy to deduce Lemma
\ref{Lemm: nbrpreimagesP'}. $\square$

\section{Estimate of $\mbE \big [ {\rm{Tr}} V_N^{s_N} \big ]$ when $\Sigma={\rm diag}(\pi
_1, 1, \ldots, 1)$} {\label{sec: expect}}

We here prove the universality of the expectation (\ref{tracemn}) in
various scales $s_N$ depending on the value of $\pi_1$ with respect to
the critical value $w_c=1+1/\sqrt \gamma$.\\
Let $c>0$ be a given real number. In the next theorem, $(s_N)$ is a sequence of integers such that 
$$\begin{cases}
\lim_{N \to \infty}\frac{s_N}{\sqrt N}=c &\text{ if $\pi_1>w_c,$} \\
\lim_{N \to \infty}\frac{s_N}{ N^{2/3}}=c &\text{ if $\pi_1\leq w_c.$}
  \end{cases}
$$

\bt \label{theo: expect} Let $V_N$ be a complex (resp. real) matrix satisfying
${\rm{(H_1)-(H_3)}}$. If  $\pi_1\geq w_c$, we also assume that $V_N$ satisfies ${\rm{(H_4)}}$ (resp. ${\rm{(H'_4)}}$).
\begin{itemize}
\item[(i)] Assume that $\pi_1 > w_c$. Then there exists a constant $\hat C_4>0$ which depends on $\max_j \mathbb E (|X_{1j}|^4)$ such that for $N$ large enough,
$$\mbE \big [
{\rm{Tr}} \, V_N^{s_N} \big ] \leq \hat C_4 \tau({\pi _1}) ^{s_N} \quad \text{
and } \quad \mbE \big [ {\rm{Tr}} \, V_N^{s_N} \big ]= \mbE \big [
{\rm{Tr}} \, (V_N^G)^{s_N} \big ] (1+o(1)).$$
\item[(ii)] Assume that $\pi_1 = w_c$. Then there exists a constant $\hat C'_4>0$ which depends on $\max_j \mathbb E (|X_{1j}|^4)$ such
that for $N$ large enough, 
$$\mbE \big [
{\rm{Tr}} \, V_N^{s_N} \big ] \leq \hat C'_4 u_+ ^{s_N} \quad \text{and} \quad \mbE
\big [ {\rm{Tr}} \, V_N^{s_N} \big ]= \mbE \big [ {\rm{Tr}} \,
(V_N^G)^{s_N} \big ] (1+o(1)).$$
\item[(iii)] Assume that $\pi_1 < w_c$. Then there exists a constant $C>0$ such
that for $N$ large enough,
$$\mbE \big [
{\rm{Tr}} \, V_N^{s_N} \big ] \leq C u_+ ^{s_N} \quad \text{ and } \quad \mbE
\big [ {\rm{Tr}} \, V_N^{s_N} \big ]= \mbE \big [ {\rm{Tr}} \,
(M_N^G)^{s_N} \big ] (1+o(1)).$$
\end{itemize}
\et

More precisely, in $(i)$ if $V_N$ is
complex, one can show that $\mathbb{E}
\left [\text{Tr} V_N^{s_N}\right]= (1+o(1))\tau(\pi _1)^{s_N}\exp \big [
({s_N^2}/{2N}) \left (\frac{\sigma{(\pi _1)}} {\tau(\pi _1)} \right
)^2 \big ] $. In the real setting, the same estimate holds
with $\sigma(\pi _1)$ replaced by $\sqrt{2} \sigma(\pi _1)$. These
estimates can trivially be deduced from Theorem \ref{theo:  BaikGBAPeche} $(i)$ and its real counterpart (due to
\cite{Debashis}) combined with some considerations close to those made in Section 2 of \cite{PecheFeral}.\\
We point out that in the three regimes, the asymptotics of $\mathbb{E} \left [\text{Tr} V_N^{s_N}\right]$ differ in the complex and real settings. This is not surprising since the limiting distributions of the largest eigenvalues are different. Through the combinatorial analysis, this fact is justified by the existence of non-MP-closed vertices in some typical paths. The investigation of such vertices is here really similar to that made in \cite{PecheWishart} and we refer to Section \ref{sec: white} above and \cite{PecheWishart} for more detail.

\paragraph{}This section is devoted to the proof of Theorem \ref{theo: expect}. The contribution to the expectation
(\ref{tracemn}) of paths with no $1$-edges has been evaluated in
Proposition \ref{Propcontriwithout1edge}. This section is devoted to
the estimation of the contribution from edge paths $P$ having 1-edges. We shall show that if $\pi _1 < w_c$, this contribution is negligible with respect to that from paths without $1$-edges (which is of the order of $\mathbb{E} \left [\text{Tr} M_N^{s_N}\right ]$, cf. Section \ref{sec: white} above). On the other hand, when $\pi _1\geq w_c$, we shall prove that paths with $1$-edges contribute in a non-negligible way. As announced, our proof will make use of the gluing procedure. Thus, we will first consider the glued
paths $P'$ and find the typical ones that is those which contribute
in a non-negligible way to the expectation. We will easily see that the typical paths $P'$ have all their edges passed twice. Then, given a typical
path $P'$, we shall estimate the contribution of all its preimages $P$. This will require to
examine the added weight due to the erased 1-edges: by construction of the gluing procedure, it may happen
that $P$ has some 1-edges passed at least four times. We shall check that the contribution of the typical paths $P$ depends only on the second and fourth moments of the $X_{ij}$'s.

\paragraph{}Before we proceed, we need a few notations. A glued path
$P'$ is of length $2(s_N-(s-l))$ where $s$ (resp. $l$) denotes the
number of pairs of 1-edges (resp. of clusters) in its preimages $P$.
Note that $s-l$ counts the number of pairs of $1$-edges that
have been erased by the gluing procedure; $l$ counts the pairs of $1$-edges in $P'$. Besides, the origin of
$P'$ is the vertex $1$ and is a non marked vertex on the bottom
line. Throughout the paper, we also denote by $m$ the number of
times the trajectory of $P'$ goes back to the level $0$. Note that
in general $m \leq l$.

\bdefi We call ${\mathbf{N}}(s_N-(s-l),k,m)$ the number of Dyck
paths of length $2(s_N-(s-l))$ with $k$ odd marked instants and $m$
returns to $0$. \edefi

The simplest case to deal with is when $m=l$, that is when all the
occurrences of the vertex $1$ on the bottom line of $P'$ are
encountered at the instants where its trajectory hits the
level $0$. Thus, when $m=l$, estimating the contribution to (\ref{tracemn}) of paths
$P$ follows from statistics on the number of
returns to $0$ of the underlying Dyck paths of $P'$ (each return is weighted by $\pi _1$). This observation justifies the following definition.

\bdefi A path $P'$ is said to be a fundamental path if all
its $1$-edges occur at level $0$. \edefi

In the following subsection, we present the detailed computations of
the contribution to (\ref{tracemn}) from edge paths
$P$ associated to a fundamental path $P'$. In Subsection \ref{subsec:
casl>m}, we consider the set of non-fundamental paths $P'$ that is
the case where some clusters in the initial paths $P$ do share edges
in such a way that some $1$-edges in $P'$ occur at levels greater
than 0. As we will see, this requires to refine the analysis and
define a new gluing procedure.

\subsection{ \label{subsec: casl=m} All the occurrences of the vertex $1$ on the bottom line are made at level $0$}
In this subsection, we investigate the set of paths $P$ such that
the $l$ different clusters $\mathcal S_j, j=1, \ldots, l$, after the
gluing procedure, yield a fundamental path $P'$. We start from such $P'$ and examine the possible added weight when reversing
the gluing process, that is the expectation from the erased 1-edges.
The simplest case, examined in the subsequent proposition, is when there is no erased $1$-edge
through the gluing process.

\paragraph{} Denote by $Z_1$ the contribution of paths $P$ having
$1$-edges passed at most twice and whose associated glued path $P'$ satisfies
$m=l=s$. $P'$ has then $s$ returns to $0$ and length $2s_N.$

\bp \label{prop: estZ1m=s} One has that
\begin{itemize}
\item[(i)] $Z_1 =O(1)\tau({\pi _1})^{s_N}$ if $\pi_1 > w_c$ and $s_N=O(\sqrt N)$,
\item[(ii)] $Z_1 =O(1)u_+^{s_N}$ if $\pi_1 = w_c$ and $s_N=O(N^{2/3})$,
\item[(iii)] $Z_1 =o(1)u_+^{s_N}$ if $\pi_1< w_c$ and $s_N=O(N^{2/3})$.
\end{itemize}
\ep

\paragraph{Proof of Proposition \ref{prop: estZ1m=s}:}By assumption, the paths $P$ and $P'$ coincide up to
a translation of the origin. Furthermore, the vertex $1$ is the
origin of $P'$, is non marked and, by definition of the gluing
procedure, the path $P'$ has $1$-edges seen at most twice. To define $P$ from $P'$, one only has to determine the origin of the
path $P$, which amounts to choosing an even instant before the first
return to $0$ of the trajectory associated to $P'.$\\
\indent Let $\alpha, \alpha'$ be such that $0<\alpha' < \frac{\sqrt \gamma}{1+\sqrt \gamma}<\alpha <1.$
Call $\hat Z_1$ the contribution from paths $P$ associated to
fundamental paths and for which the number $k$ of odd up steps
satisfies $\alpha' s_N \leq k \leq \alpha s_N$. The computations of \cite{PecheWishart} (Section 3), summarized in Section \ref{sec: white}, can be copied to show that typical paths $P'$ of length $2s_N$ having $k$ odd up steps and $s$ returns to the level 0 have
edges passed only twice. Let $\mbE_{k,s}$
denote the expectation with respect to the uniform distribution on the set of Dyck paths with $k$
odd up steps and $s$ returns to the level 0. It is in particular a minor modification to show
that the estimate (\ref{estimexpMaxPe}) holds when $\mbE_{k}$ is
replaced with $\mbE_{k,s}$ (and ${\mathbf{N}}(s_N,k)$ with
${\mathbf{N}}(s_N,k,s)$) uniformly in $s$ : the proof can be deduced
for instance using arguments given in \cite{PecheFeral},
Lemma 7.10 ($2^{nd}$ case). As already recalled in Section \ref{sec:
white}, it is also proved in \cite{PecheWishart}
that a typical path of length $2s_N$ with $s$ returns to $0$ has an
unmarked origin, no edge read more than twice and no vertex of
type strictly greater than 3. Thus, we can deduce that typical paths
$P$ in $\hat Z_1$ have edges
passed only twice. From the above, we deduce that
\begin{eqnarray} \label{est: s=m}
&\hat Z_1=&O(1) \times \sigma ^{2s_N} \sum_{1 \leq s_1\leq s_N}
\sum_{1 \leq k_1\leq s_N} s_1 {\bf{N}}(s_1-1, s_1-k_1) \sum_{m=1}^{s_N-s_1}
\crcr && \sum_{k=k_1+m -1}^{k_1+ s_N-s_1
}{\bf{N}}(s_N-s_1,k-k_1,m-1) \pi_1^m {\gamma_N^{k-s_N}}.
\end{eqnarray}
\indent We now consider the paths $P$ contributing to $Z_1$ and for which $k\geq \alpha s_N$ or $k\leq
\alpha' s_N$ and show that they contribute in a negligible way to $Z_1$. To this aim, let $\hat k:=\left [ \frac{\sqrt{\gamma}}{1+\sqrt{\gamma}}s_N \right ]+1$. Here we show that there exists a constant $C>0$ such that for any integer $n$ (with $0 <\hat k +n \leq s_N$), 
\be \mathbf{N}(s_N, \hat k +n,m) \gamma_N ^{\hat k +n} \leq C e^{-Cn^2/s_N}\mathbf{N}(s_N, \hat k ,m)\gamma_N ^{\hat k} .\label{borne sur k}\ee
This will imply that the main contribution to $Z_1$ comes from paths with approximately $\hat k$ odd marked instants, so that $Z_1=\hat Z_1(1+o(1))$. To prove $(\ref{borne sur k})$, we write 
$$\mathbf{N}(s_N, \hat k +n,m)\gamma_N ^{\hat k +n} =\sum^*_{s_1, \ldots, s_m}\sum^*_{k_1, \ldots, k_m}\prod_{i=1}^m\mathbf{N}(s_i-1, s_i-k_i)\gamma_N ^{ k_i }  ,$$
where the starred sums bear on integers $s_i$ summing to $s_N$ and $k_i$ summing to $\hat k +n.$
We also set $\hat k_i=\left [\frac{\sqrt{\gamma}}{1+\sqrt{\gamma}}s_i\right ](+1)$ so that $\sum _i \hat k _i = \hat k.$ Using the ideas of Remark 2.4 of \cite{PecheWishart}, one has that
$$\mathbf{N}(s_i-1, s_i-k_i)\gamma_N ^{k_i} \leq e^{\{-C'\frac{(k_i-\hat k_i)^2}{s_i-1}\}} \mathbf{N}(s_i-1, s_i-\hat k_i)\gamma_N^{\hat k_i}$$
for some constant $C'>0$, independently of $s_i$. Furthermore setting $k_i-\hat k_i=x_i  (s_i-1),$
one can easily show that
\begin{eqnarray*}
&&\!\!\!\!\!\!\!\!\!\sum_{k_1, \ldots, k_m}^* \exp \left ( -C' \sum_{i=1}^{m} \frac{(k_i-\hat k_i)^2}{s_i-1} \right ) \crcr
&&\!\!\!\!\!\!\!\!\!\!\!=\int\cdot  \int \exp \left ( -C' \big (\displaystyle{\sum_{i=1}^{m-1}} (s_i-1)x_i^2 
+ \frac{(n - \sum_{i=1}^{m-1} (s_i-1)x_i)^2}{s_m -1}  \big ) \right ) \prod_{i=1}^{m-1} (s_i-1) \, dx_i   \crcr
&&\!\!\!\!\!\!\!\!\!\!\!\leq  e^{-C \frac{n^2}{s_N}} \, \prod_{i=1}^{m-1} \sqrt{s_i-1},
\end{eqnarray*}
for some constant $C$ independent of the $s_i$'s and $m$. Moreover,
one can show that  $\mathbf{N}(s_N, \hat k ,m) \leq C
\prod_{i}\sqrt{s_i-1} \, \mathbf{N}(s_i-1, s_i- \hat
k_i)$. Indeed, given $s_i$, the number of $\tilde k_i$   contributing in a non negligible way  to $\sum_{\tilde k_i} \prod_{i=1}^m \mathbf{N}(s_i-1, s_i-\tilde k_i)\gamma_N ^{\tilde k_i},$ where $\sum_i \tilde k_i=\hat k$, is of the order $\sqrt {s_i-1}$ and each product term is of order of $\prod_{i=1}^m \mathbf{N}(s_i-1, s_i-\hat k_i)\gamma_N ^{\hat k_i}$ (uniformly in ${s_i}$). Thus we
get (\ref{borne sur k}) and we can conclude directly that the
contribution of paths for which $k\geq \alpha s_N$ or $k\leq \alpha'
s_N$ is negligible and that $Z_1=\hat Z_1(1+o(1))$.\\
There now remains to prove that (\ref{est: s=m})
yields Proposition \ref{prop: estZ1m=s}. This is obtained from Lemma \ref{lem: coeffsn} stated and proved below.  $\square$

\paragraph{} We now turn to the proof of the announced Lemma \ref{lem: coeffsn}.
Let $n \geq 1$ be an integer. Set  
\be \label{DefasN}
a_{n}=\sum_{s_1=1}^{n}
\sum_{k_1=1}^{n} s_1 {\bf{N}}(s_1-1, s_1-k_1) \sum_{s=1}^{n-s_1}
\sum_{k=k_1+s -1}^{k_1+ n-s_1 }
      {\bf{N}}(n-s_1,k-k_1,s-1) \frac{\pi_1^s \gamma_N^{k}}{\gamma_N^{n}},
\ee
so that $(\ref{est: s=m})=O(1) \times \sigma^{2s_N}a_{s_N}.$ 
\bl \label{lem: coeffsn}  Let $a'_n=\sigma^{2n }a_n.$ For
$n$ large enough, one has that\\
$(i)$ if $\pi_1>w_c$ then $a'_n=\tau({\pi _1})^{n}(1+o(1));$
\quad $(ii)$ if $\pi_1= w_c$ then $a'_n=u_+^{n}(1+o(1))$;\quad
$(iii)$ if $\pi_1< w_c$ then $a'_n=\frac{1}{\sqrt n} u_+^{n}(1+o(1))$.
 \el

\paragraph{Proof of Lemma \ref{lem: coeffsn}:}The proof makes use of various generating functions, for which we need a few definitions. 
Let $\chi_{n}$ denote the set of Dyck paths of length
$2n$. For a trajectory $x \in \chi_{n}$, we define
\begin{eqnarray*}&&r_x:=\sharp \{ t\in ]0, 2n[,\: x(t)=0  \}, \:\: o_x:=\sharp \{\text{ odd marked instants of $x$ }\} \crcr
 &&e_x:=\sharp \{\text{ even marked instants of $x$ }\}.
\end{eqnarray*} Introduce the generating functions
$$F( \pi_1, \gamma, z)=\pi_1 \sum_{n\geq 0}\sum_{x \in
\chi_{n}}\pi_1^{r_x}\gamma^{-e_x}z^{n},\: K(z):=\sum_{n\geq
0}(n+1)\sum_{x\in \chi_{n}}\gamma^{-o_x}z^{n+1}.$$
Then the function
\be H(z):=F( \pi_1, \gamma, z)K(z)
\ee
is ``almost'' the generating function associated to the terms $a_{n}$ and thus to those occurring
in (\ref{est: s=m}). Indeed in the definition of $a_n$ as in (\ref{est:
  s=m}), we have $\gamma_N$ instead of $\gamma$. This will have no
impact on the following reasoning as $\lim _{N\to \infty} \gamma_N= \gamma$ and we are interested in large $N$-asymptotics.
Thus expanding $H$ as a power series $H(z):=\sum_{n=0}^{\infty}a_n
 z^n$, one has that $Z_1=O(1) \times a'_{s_N}$. 

\paragraph{}In order to determine the asymptotics of $a_n$, we now turn to the evaluation of the generating functions.
This is the aim of the subsequent lemma. 
\bl
\label{lem: foncgen} One has that $$F(\pi_1,
\gamma,z)=\frac{2}{2/\pi_1-(1-\gamma^{-1})
z-1+\sqrt{(1+(1-\gamma^{-1}
 )z)^2-4z}}$$ and
$$K(z)=\frac{z}{2} \, \frac{\partial}{\partial z}\left ({(1-\gamma^{-1})z+1-\sqrt{(1+(1-\gamma^{-1})z)^2-4z}}\right).$$
\el
\paragraph{Proof of Lemma \ref{lem: foncgen}:} We need to define two auxiliary generating functions to prove Lemma \ref{lem: foncgen}.
Set $$G(\gamma, z)=\sum_{n\geq 0}\sum_{x\in
\chi_{n}}\gamma^{-o_x}z^{n}, \quad \tilde G(\gamma, z)=\sum_{n\geq
0}\sum_{x\in \chi_{n}}\gamma^{-e_x}z^{n}.$$ Then, decomposing any
trajectory $x\in \chi_{n}$ when $n>0$ according to the first return
to the origin, one deduces the following relations:
\begin{eqnarray}
&& \tilde G(\gamma, z)=1+z  G(\gamma, z)\tilde G(\gamma, z), \quad G(\gamma, z)=1+\gamma^{-1} z\tilde  G(\gamma, z) G(\gamma, z),\crcr
&& F(\pi_1, \gamma, z)=\pi_1+\pi_1z G(\gamma, z)F(\pi_1,\gamma, z).
\end{eqnarray}
Solving these equations yields (see \cite{SosWish}) that $F(\pi_1, \gamma, z)= \dfrac{\pi_1}{1-\pi_1z
G(\gamma, z)},$ where
$$ G(\gamma,
z)=\frac{(1-\gamma ^{-1})z+1-\sqrt{(1+(1-{\gamma}^{-1})z)^2-4z}}{2z}.$$ For the evaluation of $K$, it is
enough to observe that
\begin{eqnarray}
&K(z)&=z \sum_{n\geq 0}(n+1) \sum_{x \in
\chi_{n}}\gamma^{-o_x}z^{n}=z\frac{\partial }{\partial
z}\sum_{n\geq 0}\sum_{x \in \chi_{n}}\gamma^{-o_x}z^{n+1}\crcr
&&=z\frac{\partial }{\partial z}\left ( z G(\gamma, z)\right ).
\end{eqnarray}
This finally yields Lemma \ref{lem: foncgen}. $\square$

\paragraph{}Thanks to Lemma \ref{lem: foncgen}, one shall then deduce the asymptotics of $a_n$ as $n$ goes to infinity from the generating function $H(z):=F(\pi_1, \gamma, z)K(z).$
Set $U=zG(\gamma, z).$
It can be deduced from \cite{SosWish} (pp. 21-22) that $U$ is
holomorphic in the disk $\{z, |z|<\sigma^2/u_+\}$, and one has that
$z=\frac{U(U-1)}{(1-\gamma^{-1})U  -1}.$ Furthermore, $z=0$ if
$U=0.$
Assume first that $\pi_1\leq w_c.$ One thus has that
$$a_n:=a'_n/\sigma ^{2n}=\frac{1}{2i\pi}\int_{\mathcal{C}_o}\frac{1}{z^n} \, \frac{\frac{d}{dz}(zG(\gamma,z))}{1/\pi_1-zG(\gamma, z)} \, dz,$$
where the contour $\mathcal{C}_o$ encircles $0$, is oriented counterclockwise and lies in the disk $\{z, |z|<\sigma^2/u_+\}$.
By a straightforward change of variables, one gets that
$$a_n=\frac{1}{2i\pi}\int_{\mathcal{C}} \frac{1}{1/\pi_1-u}\left (\frac{(1-\gamma^{-1})u-1}{u(u-1)} \right)^n \, du,$$
where $\mathcal C$ is a symmetric contour encircling $0$ and
remaining on the left of ${\sigma}/{\sqrt{u_+}}.$ Note that this
implies that the contour $\mathcal C$ cannot encircle $1/\pi_1.$ It
is then an easy saddle point argument to check points $(ii)$ and
$(iii)$: the critical point is $u_c:={1}/{w_c }$ and the saddle
point contour is modified in a neighborhood of width $1/\sqrt n$ of
$u_c$ so that $\mathcal C$ remains to the left of $u_c.$\\
When $\pi_1>w_c$, the contour $\mathcal{C}$ does not
encircle $1/\pi_1$ and a straightforward Laplace method leads to point $(i)$. This finishes the proof of Lemma \ref{lem: coeffsn}.
$\square$

\paragraph{}We now turn to estimating the contribution of paths $P$ with $1$-edges seen only twice and which give a fundamental glued
path $P'$ by erasing a positive number of $1$-edges (precisely $2(s-l)$ with our
notations). We call $Z_2$ this contribution. Note that the
glued paths $P'$ to be considered here are such that $m=l$ with $m<s$.

\bp \label{Prop: Z2} There exists a constant $C>0$ such that
$Z_2\leq CZ_1.$
\ep

The proof will make use of the following extension of Lemma
\ref{lem: coeffsn}. In the next lemma, we write $a'_{n}=a'_{n}[\pi
_1]$ (recall that $a_n'=\sigma^{2n } a _n$ with $a_n$ given by (\ref{DefasN})).

\bl \label{lem: coeffsnbis} Assume that $\pi_1 \leq w_c.$ Let $C>0$ be some constant independent of $N$. Then for all
large $N$, and as long as $s_N=O(N^{2/3}),$ there exists a constant
$C'$ depending on $C$ only such that
$ a'_{s_N} \big[  \pi _1 e^{{Cs_N}/{N}} \big ]=a'_{s_N}[\pi _1](C'+o(1)).$
\el We skip the proof of this lemma which can be obtained by the
same saddle point argument as in Lemma \ref{lem: coeffsn}. We now
turn to the proof of Proposition \ref{Prop: Z2}.

\paragraph{Proof of Proposition \ref{Prop: Z2}:}Let $P'$ be a fundamental path. It can first be deduced from Lemma
\ref{Lemm: nbrpreimagesP'} (applied with $l=m$) that the number of
paths $P$ which are preimages of $P'$ and
contribute to $Z_2$ is at most \be \label{EstimNbrpreimagesP'Z2}
s_1 \frac{ ( 2ss_N )^{s-m}}{(s-m)!}\ee where $s_1$ is such
that the first return to zero of $P'$ occurs at time $2s_1.$ We already know from
\cite{PecheWishart} and \cite{Sos} that in typical paths $P'$ no
edge is read more than twice. 
As it is assumed that $P$ has no $1$-edges seen $4$ times or more, the
expectation of $P$ is just $(\pi_1\sigma^2/p)^{s-m}$ times that of $P'$. Using
the inequality ${\bf{N}}(s_N-(s-m),k,m-1)\leq {\bf{N}}(s_N,
k+s-m,s-1)$ and setting $s'=s-m \geq 1$ and $k'=k+s'$, one can check
that there exists a constant $C>0$ (whose value may vary from line to
line) such that
\begin{eqnarray}
Z_2 & \leq & C \sigma ^{2s_N} \sum_{s_1=1}^{s_N}\sum_{s=1}^{s_N-s_1}\sum_{m=1}^{s-1}\sum_{k=1}^{s_N} \sum_{k_1\leq k'}
  s_1 {\bf{N}}(s_1-1,
s_1-k_1)\crcr && {\bf{N}}(s_N-(s-m)-s_1,
k-k_1,m-1)\frac{\left (\frac{2ss_N}{p} \right)^{s-m}}{(s-m)!} \,
{\pi_1^s } \, {\gamma_N^{k-s_N}} \label{Z2<<Z1}\\ 
& \leq & C \sigma ^{2s_N}\sum_{s_1=1}^{s_N}\sum_{s=1}^{s_N-s_1}
  \sum_{s'=1}^{s_N}\sum_{k'=s'+1}^{s_N}\sum_{k_1\leq
k'}s_1{\bf{N}}(s_1-1, s_1-k_1)\crcr
&& {\bf{N}}(s_N-s_1, k'-k_1,s-1)\frac{1}{s'!}\left (\frac{2\gamma_N^{-2} ss_N}{N} \right)^{s'} \, {\pi_1^s } \, {\gamma_N^{k'-s_N}}  \crcr
&\leq & C \sigma ^{2s_N}  \sum_{s_1=1}^{s_N}\sum_{s=1}^{s_N-s_1 }\sum_{k'=1}^{s_N}
 \sum_{k_1\leq k'}s_1{\bf{N}}(s_1-1,
s_1-k_1)\crcr && {\bf{N}}(s_N-s_1, k'-k_1,s-1) \big{\{} \exp \left
({C ss_N}/{N} \right) -1 \big{\}} \, {\pi_1^s } \,
{\gamma_N^{k'-s_N}} \label{Z2<<Z1bis}.
\end{eqnarray}
In (\ref{Z2<<Z1}), the factor $\left ({2 ss_N}/{p}\right)^{s'}$ 
can be deduced from (\ref{EstimNbrpreimagesP'Z2}) and the fact that the 
rescaling factor $p^{s_N}$ splits into $p^{s_N-s'}p^{s'}.$ 
In the case where $\pi _1> w_c$ and $s_N=O(\sqrt N)$, we then readily get the
result. For the case where $\pi _1 \leq w_c$ and $s_N=O(N^{2/3})$, the
conclusion follows from Lemma \ref{lem: coeffsnbis}. $\square$

\paragraph{}Amongst the paths $P$ associated to a fundamental glued path $P'$, there remains to consider those with some $1$-edges seen
four times or more. By definition of the gluing procedure, these edges
must be erased and appear at most twice in $P'$. Thus, as for $Z_2$,
one has that $m=l$ and $s>m$ (and there are at most $(s-m)$ $1$-edges
seen 4 times or more in $P$). We call $Z_3$ the contribution of these
paths and we show that those contributing to $Z_3$ in a non-negligible
way have $1$-edges passed at most 4 times.

\bp \label{Prop: Z3} There exists a constant $\hat C_4>0$ which depends on the fourth moments of the entries of $X$ such that $Z_3 \leq \hat C _4 Z_2$. \ep

Observe that $Z_3$ is non-negligible when $\pi _1\geq w_c$ 
which partly explains the added constraint on the fourth moments of
the $X_{ij}$'s to get the announced universality in cases $\pi_1>w_c$ and $\pi_1=w_c$.

\paragraph{Proof of Proposition \ref{Prop: Z3}:} One already knows from \cite{PecheWishart} and \cite{Sos} that in
typical paths $P'$ no edge is read more than twice. Yet, in paths
contributing to $Z_3$, there exist some vertices that occur more
than twice on each of the top and bottom lines. Thus choosing $s-m$
moments of time in $P'$ (to reconstruct $P$) can result into a
$1$-edge which is read more than twice in $P$. Note that by definition of clusters, such an edge can only be
read inside one
cluster in $P$. We now estimate the
expectation of the path $P$ with respect to that of $P'$. We call $\hat Z_3$ the contribution to $Z_3$ from paths $P$ with 1-edges read 4 times at most and $Z_3^0$ denotes the remaining contribution to $Z_3$.
We show that $\hat Z_3$ is of the order of $Z_2$ while $Z_3^0=o(1) Z_2$.
Our reasoning is mainly based on several properties of the gluing
procedure which has been defined in Section \ref{sec: counting} above.\\
Consider a $1$-edge $e=\begin{pmatrix}v\\
1 \end{pmatrix}$ which is read at least $4$ times in $P$. Assume it is
read $2y_1 \geq 4$ times in $P$ which, using $({\rm H} _2)$, implies that its expectation is at most
$(Cy_1)^{y_1}$ for
some constant $C>0$ independent of $N$.
%
\paragraph{} {\it $1^{rst}$ case:} The edge $e$ does not coincide with any of the edges of $P'$. In other words, $e$ is distinct from the edges starting the different clusters in $P'$. 
This means that amongst the $s-m$ instants in $P'$
where we have erased $1$-edges (that is the set $\mathcal T$ in Section \ref{sec: counting}), we
have chosen $y_1$ times edges with vertex $v$ on the top line. These
choices split the subpath $P_1^g$ (derived from the cluster
$\mathcal{S}_1$ by the gluing procedure) into $y_1+1$ subpaths as
follows:
\begin{equation}{\label{decompy1}}
\begin{pmatrix}g_s\\ 1 \end{pmatrix}
\cdots  \begin{pmatrix}v\\ \alpha_1
\end{pmatrix}\begin{pmatrix}v\\ \beta_1 \end{pmatrix}\cdots
\begin{pmatrix}v\\ \alpha_2 \end{pmatrix}\begin{pmatrix}v\\ \beta_2 \end{pmatrix}\cdots \begin{pmatrix}v\\ \alpha_{y_1} \end{pmatrix}\begin{pmatrix}v\\ \beta_{y_1} \end{pmatrix}\cdots \begin{pmatrix}g_s\\ 1 \end{pmatrix}.
\end{equation}
Denote now by $\tilde Q_j, j =1, \ldots,
y_1-1,$ the subpath starting with the edge $\begin{pmatrix}v\\
\beta_j \end{pmatrix}$ and ending with the edge $\begin{pmatrix}v\\
\alpha_{j+1} \end{pmatrix}.$ Let also $\tilde Q_o$ (resp. $\tilde
Q_f$) be the subpath starting with $\begin{pmatrix}g_s\\ 1
\end{pmatrix}$ (resp. $\begin{pmatrix}v\\ \beta_{y_1}
\end{pmatrix}$) and ending with
$\begin{pmatrix}v\\ \alpha_1 \end{pmatrix}$ (resp. $\begin{pmatrix} g_s\\
1 \end{pmatrix}$). Then each path $\tilde P'$ which is obtained from
$P'$ by permuting any of the $\tilde Q_j, j=1, \ldots , y_1-1$ leads
by permuting the paths $P_i$ to the same path $P$. Thus, the number
of preimages of such a path $P'$ has to be divided by a factor
$(y_1-1)!$ since each preimage is counted $(y_1-1)!$ times when
considering all the possible paths $P'$ (recall the proof of Lemma
\ref{Lemm: nbrpreimagesP'}). Taking into account the expectation
of the edge $e$ in $P$ then adds a factor $(Cy_1)^{y_1}/(y_1-1)! \leq C^{y_1}$.\\
Let us count now the number of ways to select $s-m$ moments of time in such a way that we define $y_1$ times the same edge $e$. For this, assume that the instant $\tilde t $ where
$\tilde Q_1$ begins in $P'$ has been chosen. Then two situations may
happen when choosing the instant $\tilde t'$ where it ends. To
explain this, we need to introduce two characteristics (already mentioned in
Section \ref{sec: white}) of the path $P'$. The first one is $\nu
_N=\nu _N(P')$, the maximal number of vertices that can be visited in $P'$
at marked instants from a given vertex different from the origin $1$. The second
one is $T_N=T_N(P')$, the maximal type of a vertex in $P'$. We
shall use the following fact deduced from the very
definitions of $\nu _N$ and $T_N$ : given a vertex (different from the origin $1$) occuring in $P'$,
it appears at most $T_N+\nu_N$ (resp. $T_N$) times as endpoint
(resp. right endpoint) of up steps. 
It is then not hard to see that a given vertex (distinct of the origin $1$) appears at most 
$2(T_N+\nu_N)$ times along the path $P'.$ 
Thus, the number of ways to select $y_1$ times the same vertex $v$
when choosing (in $P'$) the $s-m$ moments of time does not exceed:
\begin{eqnarray*}
&&\binom{s_N}{s-m-y_1} \times (s_N-s+m+y_1) \times \binom{2(T_N+ \nu_N)}{y_1-1}\crcr
&& \leq C^{s-m}\binom{s_N}{s-m}\left ( \frac{2(T_N+\nu_N)}{s_N-s+m}\right)^{y_1-1},
\end{eqnarray*}
for some constant $C>0.$

\paragraph{} {\it $2^{nd}$ case:} The edge $e$ coincides with one of the edges of $P'$. 
In this case, the above reasoning on the permutation of the subpaths $\tilde Q_j$ still applies.
One then needs to distinguish two cases according to the value of $y_1$. \\
\indent Case $(a)$: $y_1=2.$ Then the edge $e$ is seen exactly four times in $P.$
To determine $e$ one has to select one of the $m$ edges starting a
cluster. This determines the vertex $v$. The occurrence of $v$ along
the path $P'$ where we have erased $e$ has then to be determined. In principle, there are at
most $\nu_N+T_N$ possible choices for this occurrence. Nevertheless it
is the most probable that $m(T_N +\nu_N)>>s_N$ due to the fact that
$m$ may be large. Thus, calling on the characteristics $\nu _N$ and
$T_N$ does not improve the estimate and it is sufficient to notice, as
in Lemma \ref{Lemm: nbrpreimagesP'}, that the number of ways to choose the $s-m$ moments of times to determine the erased $1$-edges is at most of order 
$$  \binom{s_N}{s-m},$$
that is exactly as for $Z_2$.\\
\indent Case $(b)$: $y_1>2$. In this case, once $v$ is determined (with at most $s_N$ ways to do so), there are at most $\binom{2(T_N+\nu_N)}{y_1-2}$ possible ways to select $y_1-2$ other repetitions of $v$ in $P'$. 
Thus, the number of ways to select the $s-m $ moments of time to determine the $1$-edges in this case does not exceed
$$C^{s-m}\binom{s_N}{s-m}\left ( \frac{2(T_N+\nu_N)}{s_N-s+m}\right)^{y_1-2}.$$

We can now conclude that $Z_3^0$ is negligible. Indeed, we deduce that
\begin{equation} \label{estimweight}
Z_3^0 \leq 2 Z_2 \sum _{y_1 \geq 3}  \left(\frac{2C'(T_N+\nu_N)}{s_N-s+m} \right ) ^{y_1 -2}=o(1)Z_2,
\end{equation}
since for typical paths $P'$, one can show that $s_N-s+m\to \infty$ (using (\ref{Z2<<Z1}) and the above) and
that $(T_N+ \nu_N)^2<<s_N-s+m$ (this point, recalled in Section
\ref{sec: white} follows from \cite{PecheWishart}, Section 3.2). 
Similarly we can also show that the contribution of paths $P$ with
$1$-edges seen $4$ times in $P$ but that do not arise in $P'$ (this
corresponds to the situation of the previous {\it $1^{rst}$ case} with
$y_1=2$) is negligible with respect to $Z_2$ and does not contribute
to the expectation (\ref{tracemn}).\\
Last, it is not hard to see from Case $(a)$ that $\hat Z_3$ is of the
order of $Z_2$. More precisely, for each edge $e$ seen four times in
$P$ and twice in $P'$, one has to multiply the expectation of $P'$ by ${\pi_1\mbE |X_{1v}|^4}/{(\sigma^2 p)}$ at most to get that of $P$. In the sequel we set $\hat C _4 :=\max_v \mathbb E |X_{1v}|^4/ \sigma ^2+1$.
As $s-m$ counts the number of instants in $P$ where a $1$-edge has been erased, one has that (compare with (\ref{Z2<<Z1}))
\begin{eqnarray}
&\hat Z _3\leq  & C \sigma ^{2s_N} \sum_{s_1=1}^{s_N}\sum_{s=1}^{s_N-s_1} 
\sum_{m=1}^{s-1} \sum_{k=1}^{s_N} \sum_{k_1\leq k+s-m} s_1 {\bf{N}}(s_1-1,
s_1-k_1)\crcr && {\bf{N}}(s_N-(s-m)-s_1,
k-k_1,m-1)\frac{\left (\frac{2\hat C_4ss_N}{p} \right)^{s-m}}{(s-m)!}
\pi_1^s  \frac{\gamma_N^{k}}{\gamma_N^{s_N}}.
\end{eqnarray}
We then conclude (using Lemma \ref{lem: coeffsnbis} in the case where $\pi _1\leq w_c$) that there is another constant $\hat C' _{4}$ depending on the fourth moments of the $X_{ij}$'s such that $\hat Z _3 \leq \hat C' _{4} Z_2$. 
This finishes the proof that typical paths $P$ may have edges seen $4$ times but not more. Note that this happens when $\pi_1\geq w_c$ only, and in this case their associated path $P'$ has no ($1$-)edge seen more than twice. $\square$ 

\subsection{Edges shared by clusters \label{subsec: casl>m}}
In this section we investigate paths $P$ such that the clusters do
share some edges in such a way that the $1$-edges in the glued path
$P'$ are not necessarily read at moments of time where the
trajectory goes back to the level $0$. Keeping the same notations as before, the length of the path $P'$ is now $2(s_N-(s-l))$ and $l-m$ returns to
the vertex $1$ on the bottom line of $P'$ occur at some positive levels. To consider such paths, we define a second gluing procedure and associate a second path $P''$ to the initial path $P$. This gluing
procedure is close to the {\it construction procedure}
already used in \cite{Si-So2} and \cite{Sos}.

\paragraph{}For short, we call $Q_i$ instead of $P_i^g$ the subpaths in-between
two returns to the vertex $1$ on the bottom line of $P'$ (recall
Section \ref{sec: counting}). We let $i_1\leq l-m-1$ be the smallest
index where the first return to the vertex $1$ on the bottom line
occurs at some positive level. Then there exists an edge which is
opened but not closed in
$Q_{i_1}$. We denote by $\hat e=\begin{pmatrix}\alpha \\
\beta\end{pmatrix}$ the first of these edges. When reading $P'$, let
then $i_2$ be the lowest index such that the edge $\hat e$ is closed (and
odd) in $Q_{i_2}$. Let then $e$ be the first edge in $Q_{i_1}$ occuring also in $Q_{i_2}$. Note that it may happen that $e\not=\hat e$: this arises in non typical paths only, as this implies that $P'$ has edges seen at least $4$ times. Let also $t_e$ and $t'_e$ be the instants of the
first occurrence of the edge $e$ in $Q_{i_1}$ and $Q_{i_2}$ respectively. We then
define the path $Q_{i_1}\vee Q_{i_2}$ obtained by the gluing of the
two subpaths by erasing the first occurrence of the common edge $e$
in each of the subpaths as follows. We first read $Q_{i_1}$ until
the left endpoint of the edge $e$ at time $t_e$. Then we switch to
$Q_{i_2}$ in the following way. If $t_e$ and $t'_e$ are of the same
parity, we then read $Q_{i_2}$, starting from $t'_e$, in the reverse
direction to the origin and restart from the end of $Q_{i_2}$ until
we come back to the instant $t'_e + 1$. If $t_e$ and $t'_e$ are not
of the same parity, we read the edges of $Q_{i_2}$ in the usual
direction starting from $t'_e + 1$ and until we come back to the
instant $t'_e$. We have then read all the edges of $Q_{i_2}$ except
the edge $e$ occurring between $t'_e$ and $t'_e + 1$. We then read
the end of $Q_{i_1}$, starting from $t_e + 1$. Having done so, we
obtain a path $Q_{i_1}\vee Q_{i_2}$ which has the same final (and
first) edge as $Q_{i_1}$ and the vertex $1$ is marked once on the
bottom line. We then set $P'_1$ to be the path defined by
$$P'_1=Q_1 \cup  \ldots \cup Q_{i_1-1}\cup Q_{i_1}\vee Q_{i_2} \cup \hat Q_{i_2} \ldots \cup Q_{i_m}.$$
Here the hat means that the corresponding term does not appear. We
then replace $P'$ with $P'_1$ and restart the same procedure. We
call $1 \leq g \leq l-m$ the number of gluings needed so that all
the occurrences of $1$-edges correspond either to a marked instant
or to a return to $0$ of the associated trajectory. Note that $g$ is
defined by $l-m-g= \sharp \mathcal D$ where $\mathcal D$
is the set of the subpaths $Q_i$ which are sub-Dyck paths of origin
1 with all their edges even and which occur at some positive levels in all the
successive $P_i'$'s. Note that through the gluing process, such
subpaths are not modified but are moved to the level 0 in the order
they appear. We denote by $P''$ the path finally obtained after $g$ steps of the
gluing procedure. By definition of this gluing procedure, $P''$ is of
length $2(s_N-(s-l)-g)$ with exactly $m'=l-g$ returns to the level
$0$, its origin is the vertex $1$ and is marked $l-m'(=g)$ times on the bottom
line. In the following, we denote by $k$ the number of odd marked
instants in $P''$.

\paragraph{}We shall now estimate the number of preimages $P$ of such a path $P''$ as well as their expectation. The first and main work here is to investigate the step from $P''$ to $P'$. Once this is done, it will be quite straightforward to estimate
the number of preimages $P$ of such a path $P'$ and their expectation
by extending the analysis made in the previous subsection. To
reconstruct $P'$ from $P''$, we have to "recover" each of the $g$
glued subpaths $Q_{i_2}$. Thus, for each glued subpath $Q_{i_1}\vee
Q_{i_2}$, we have to find the instants where $Q_{i_2}$ begins and
ends (that is the two instants of switch from one path to the
other); we also need to determine the direction in which $Q_{i_2}$
is read as well as the origin of $Q_{i_2}$ in $P''$. Actually, the origin of
$Q_{i_2}$ is just given by the marked occurrence of the vertex 1 in
$Q_{i_1}\vee Q_{i_2}$. Then, we shall take into account the weighted
contribution to the expectation of each erased edge $e$. More
precisely, one already knows that in typical paths $P''$, each edge
appears only twice. But when rebuilding $P'$ from $P''$, it may
happen that some of the erased edges $e$ appear more than twice in
$P'$. As we will see, such paths will lead to a negligible
contribution to the expectation (\ref{tracemn}), which will ensure
the universality.

\paragraph{}Consider paths $P$ having some non
disjoint clusters so that some $1$-edges in the associated glued path
$P'$ occur at positive levels. Denote by $Z_4$ the contribution to the expectation (\ref{tracemn}) from such paths $P$ for which all the erased edges $e$ between
$P'$ and $P''$ appear exactly twice in $P$ (or $P'$).

\bp \label{prop: estZ1'm not=s} The main contribution to $Z_4$
comes from paths $P$ with all edges seen twice except $1$-edges which possibly occur 4 times. And there exists a constant $\hat C_4>0$ depending on the fourth moments of the entries of $X$ such that $Z_4 \leq \hat C_4 Z_1$. \ep

\paragraph{Proof of Proposition \ref{prop: estZ1'm not=s}:}
Here we only consider paths $P''$ having all their edges seen twice
since, as previously said, these are the typical paths.

Let us first reconstruct $P'$ from $P''$. Due to the fact that the
vertex $1$ is marked $g=l-m'$ times on the bottom line of $P''$, the
weighted number of such paths $P''$ is at most of order
$$\binom{s_N-(s-l)-g-k}{g} \left ({C}/{N}\right )^{g}$$
times the weighted number of paths where the origin $1$ is non marked and with the same length, the same
number of odd marked instants and the same
number of returns to $0$. Then, by
definition of the gluing procedure, the first step beginning the
subpath $Q_{i_2}$ in $Q_{i_1} \vee Q_{i_2}$ is up which implies that
the total number of ways to choose the instants of initial switch
is at most of $\binom{s_N-(s-l)}{g}$. Moreover, the number of
possible choices for the instants where one switches for the second time from the
subpaths $Q_{i_2}$ to the $Q_{i_1}$ is at most
$\binom{s_N-(s-l)-g}{g}g!$ (the factor $g!$ comes from the fact that the $Q_{i_2}$ may be interlaced). It remains to choose the direction and the
order in which the subpaths $Q_{i}$ are read. We claim that this yields a factor $2^g
\binom{m'+g}{g} \times g! $. Indeed, one can notice that once the
$Q_{i_2}$'s are identified, the remaining $m'+g$ subpaths are known
i.e. one knows the $Q_{i_1}$'s and the paths belonging to the set $\mathcal D$. Moreover, by construction of the gluing process, these latter subpaths appear in the same relative order as in $P'$. To reorder the $Q_i$'s, one needs first to choose the place where one actually encounters the (unordered) subpaths that are glued and moved by the gluing process i.e. the $Q_{i_2}$ (note that this also reorders those belonging to the set $\mathcal D$). There are at most $\binom{m'+g}{g}$ ways to do this. Last, the previous term $g!$ counts the number of ways to reorder the subpaths
$Q_{i_2}$. Regarding the respective weights of the paths $P''$ and
$P'$, one has to take into account the erased edges. As we assume
that the erased edges are pairwise distinct and read exactly twice
in $P$, the weight of $P'$ is of order $N^{-g}$ that of $P''$.

Now, one has to reconstruct $P$ from $P'$. In fact, this is really
close to the analysis made in the previous Section \ref{subsec: casl=m}. Indeed, the upper bound on the number of preimages $P$ of a
path $P'$ obtained in Lemma \ref{Lemm: nbrpreimagesP'} does not use
the assumption that clusters are disjoint or not. Thus we deduce from Lemma \ref{Lemm:
nbrpreimagesP'} that the number of preimages $P$ of a path $P'$ is
at most $s_1 \, \binom{s}{l} \left ( 2s_N\right )^{s-l}$, if the first
return to 0 of the trajectory of $P'$ holds at time $2 s_1$.

We are now in position to estimate the contribution $Z_4$. Observe
first that 
${\bf{N}}(s_N-s_1-(s-l)-g,k-k_1,m'-1) \leq {\bf{N}}(s_N-s_1,k+(s-l)+g-k_1, s-1)$ (recall that
$m'+s-l+g=s$). Hence, letting $k'=k+(s-l)+g$ and using the
fact that $m'=l-g$, by computations similar to those made for the $Z_i$, $i=1,2,3$ in the preceding section, one has that
\begin{eqnarray}
Z_4 &\leq & C 
 \sum_{s_1=1}^{s_N} \sum_{s=1}^{s_N-s_1}
\sum_{l=1}^{s} \sum_{g=0}^l \sum_{k', k_1} s_1{\bf{N}}(s_1-1,
s_1-k_1){\bf{N}}(s_N-s_1 ,k'-k_1,s-1)  \crcr & & \frac{1} {g!}\bigg
( \frac{s_N ^2}{N} \bigg ) ^g \binom{l}{g} \bigg ( \frac{2\gamma_N
^{-1}s_N }{N} \bigg ) ^g \binom{s}{l} \bigg ( \frac{2 \gamma_N ^{-2} \hat C_4 
s_N }{N} \bigg ) ^{s-l}  \pi _1 ^s \, \gamma_N ^{k'-s_N}\sigma ^{2s_N}  \label{est1: sNot=m} \\
&\leq &  C \sigma ^{2s_N}  \sum_{s_1=1}^{s_N} \sum_{s=1}^{s_N-s_1}
\sum_{g,k', k_1} s_1{\bf{N}}(s_1-1, s_1-k_1){\bf{N}}(s_N-s_1
,k'-k_1,s-1) \crcr && \frac{1} {g!}\bigg ( \frac{s_N ^2}{N} \bigg )
^g  \exp \bigg( \frac{\hat C'_4ss_N}{N} \bigg )  \pi _1 ^s \, \gamma_N
^{k'-s_N} \label{est12: sNot=m}
\end{eqnarray}
where we used the fact that $l \leq s$. $C$ and $\hat C_4 $ are positive constants independent of $N$ but $\hat C_4 $ depends on the fourth moments of the
entries of $X$; $\hat C'_4>0$ is another constant depending on 
$\hat C_4$. In the case where $s_N=O(\sqrt N)$, one can readily see that
\begin{eqnarray*}
(\ref{est12: sNot=m}) & \leq & C \sigma ^{2s_N} \exp (
\frac{(\hat C'_4+1)s_N^2}{N} ) \sum_{s_1=1}^{s_N} \sum_{s=1}^{s_N-s_1}
\sum_{k', k_1} s_1{\bf{N}}(s_1-1, s_1-k_1)\crcr &&{\bf{N}}(s_N-s_1
,k'-k_1,s-1) \pi _1 ^s \, \gamma_N ^{k'-s_N} 
\crcr & = & \hat C'_4 \, Z_1.
\end{eqnarray*}
In the scale $s_N \sim N^{2/3}$, the above estimate needs to be
refined (since $s_N^2 >> N$). In fact, we can improve the bound on
the number of choices of the starting instants of the subpaths
$Q_{i_2}$. To see this, call 
$t'_1<t'_2<\cdots <t'_{g}$ the instants corresponding to the end of the $Q_{i_2}$'s.
Denote by $t_1<t_2<\cdots<t_{g}$ the instants beginning the reading of the
$Q_{i_2}'s$. By definition of the gluing process, each edge started at the instant $t_i$ is an up edge. Furthermore, if 
$x''$ denotes the Dyck path associated to the path $P''$, one has for
any $i=1, \ldots, g$ that
$x''(t)\geq x''(t_i)>0, \forall t\in [t_i, t'_i]$. Thus, the interval $[t_i, t'_i]$ is included in one sub-Dyck path of $x''$. 
We claim that the total number of ways to choose $t_i$ and $t'_i$ does not exceed $Cs_N^{3/2}$ for some constant $C>0$. 
Indeed, choosing $t'_i$ determines the sub-Dyck path of $x''$ containing 
$[t_i, t'_i]$. We call $X_j$ this sub-Dyck path, $2L_j$ its length and
$k_j$ the number of its odd up steps. Our estimate is obvious in the case where $L_j \leq s_N^{1/2}$.
If $L_j\geq s_N^{1/2}$, we call $N(t'_i)$ the number of ways to determine $t_i$.
Let $\mathbb E_{L_j,k_j}$ denote the expectation with respect to the
uniform distribution on the set $\chi_{L_j, k_j}$ of Dyck paths of
length $2L_j$ with $k_j$ odd up steps. Then there exists some constant $C>0$ independent of $N$, $k_j$ and $L_j$ such that (for typical $k_j$'s) 
\begin{equation}{\label{estENte1}}
\mathbb E _{L_j,k_j} \left ({N(t'_i)} / {s_N^{1/2}} \right) \leq C.
\end{equation}
The above bound essentially follows from the estimation obtained in Section 2.5 in \cite{PecheWishart}. Indeed, setting $T_{0,n,k}:= \# \chi_{n,k}$ for any $n,k$, one has that 
$$ \mathbb E _{L_j,k_j} \left ( {N(t'_i)} \right) \leq \sum _{n,k'} \frac{4 \inf \{n,(L_j - n)\} T_{0,n,k'} T_{0,L_j - n, k_j-k'}}{T_{0,L_j,k_j}} \leq C s_N^{1/2}.$$
The term $4\inf \{n, (L_j - n)\}$ counts the number of ways to choose $t_i$ once given $t'_i$ and $2n$ which is the length of the sub-Dyck paths between $t_i$ and the first return to $x(t_i)$ followed by a down step.
Given $g>0$ and a Dyck path $X$ of length $2L$, we set $$K_N^{\otimes g}(X):=\sum_{1\leq t'_1<t'_2<\cdots <t'_g\leq 2L}\prod_{i=1}^g N(t'_i),$$
where the sum bears on $t'_i$ such that $X(t'_i)>0, \forall i=1, \ldots ,g.$
Similarly and using the Appendix in \cite{PecheSos1}, 
one can show that 
there exists a constant $C>0$ independent of $N$, $k_j$ and $L_j$ such that (for typical $k_j$'s) 
\begin{equation}\label{estimexpMaxReturn2}
\mbE _{k_j,L_j} \Bigl[ K_N^{\otimes g}(X_j) \Bigr ] \leq \left (C s_N^{3/2}\right )^g.
\end{equation}

In the sequel, we call $\chi_{s_N-s_1, k'-k_1, s-1}$  the set of Dyck paths
$\tilde x$ in $\chi_{ s_N-s_1, k'-k_1}$ with $s-1$ returns to $0$. Let $\mbE_{k',s,  s_N} $ denote the expectation with respect to the uniform distribution on Dyck paths of length $2  s_N$ with $s$
returns to the level $0$ and $k'$ odd marked instants.
Let $x_1$ be the Dyck path defining the first return to $0$ of $x''$. The estimate (\ref{est1: sNot=m}) may then be replaced by
\begin{eqnarray}
&&  \sigma ^{2s_N}
 \sum_{s_1=1}^{s_N} \sum_{s=1}^{s_N-s_1}
\sum_{l=1}^{s} \sum_{g=0}^l \sum_{k', k_1} \:\sum_{x_1 \in
  \chi_{s_1-1,s_1-k_1}}s_1\: \frac{\gamma_N ^{k'}}{\gamma_N^{s_N}}\sum_{\tilde x \in \chi_{ s_N-s_1, k'-k_1,
    s-1}} \frac{C}{g!} \crcr & & 
\mbE_{k',s, s_N}\bigg[ \bigg( \frac{K_N^{\otimes g}(x_1\cup \tilde x)  }{N}
\bigg ) ^g \bigg] \binom{l}{g} \bigg ( \frac{2s_N }{\gamma_N N} \bigg ) ^g \binom{s}{l} \bigg ( \frac{2  \hat C_4 
s_N }{\gamma_N^2 N} \bigg ) ^{s-l}  \pi _1 ^s . \label{pourZ5}
\end{eqnarray}
We then deduce that there is a constant $\hat C'_4>0$ (depending on $\hat C_4$) such that $
(\ref{pourZ5}) \leq  \hat C'_4  Z_1.$
This ends the proof of Proposition \ref{prop: estZ1'm not=s}. $\square$

\brem Note that we have also shown that paths $P'$ having at least
one return to the vertex $1$ on the bottom line at some positive level lead to a negligible contribution in any scale
$s_N<<N^{2/3}$. This follows from (\ref{estimexpMaxReturn2}). \erem


\paragraph{}To complete the analysis, there remains to consider paths $P$ leading to a glued path $P'$ having some $1$-edges which occur at positive levels and for which some of
the erased edges $e$ in-between $P'$ and $P''$ appear 4 times or more
in $P$ (or in $P'$). We call $Z_5$ their contribution to the expectation (\ref{tracemn}).

\bp \label{prop: estZ2'm not=s} One has that $Z_5=o(1) Z_1.$\ep
\paragraph{Proof of Proposition \ref{prop: estZ2'm not=s}:}
Consider the set of paths $P$ such that the edges which are erased
in-between $P'$ and $P''$ and which arise at least four times in $P$
do not appear in $P''$. We call $Z_5^0$ their contribution to $Z_5$ and set $\hat Z_5:=Z_5-Z_5^0.$ 
\paragraph{}
We first show that $Z_5^0<<Z_1$. Let $e=\begin{pmatrix}\alpha \\
\beta\end{pmatrix}$ be one of the erased edges in-between $P'$ and $P''$ which appears 4 times or more
in $P'$ (or in $P$). Denote by $n_e>1$ the number
of times where $e$ is an erased edge in the gluing process. We here assume that $P$ contributes to $Z_5^0$ so that $e$ appears exactly $2n_e$ times in $P$. Thus there
are $n_e$ pairs of paths $(Q_{i_1}, Q_{i_2})$ in $P'$ such that the
derived glued subpath $Q_{i_1} \vee Q_{i_2}$ is associated to the
same edge $e$. \\
Let us first prove that this decreases the number of
preimages of the path $P''$ by a factor of
\begin{equation}\label{estZ'2}
\left ( \frac{4C(\nu _N + T_N)}{s_N} \right ) ^{n_e -1}
\end{equation}
where the quantities $\nu _N= \nu _N(P'')$ and $T_N=T_N(P'')$ have
already been defined and used in the proof of Proposition \ref{Prop:
Z3}. For this, consider the first (resp. second) of the previous
considered pairs $(Q_{i_1}, Q_{i_2})$ and denote by $t_{e,1}$ (resp.
$t_{e,2}$) the instant where $Q_{i_2}$  begins and by $t'_{e,1}$
(resp. $t'_{e,2}$) the instant where $Q_{i_2}$ ends in $Q_{i_1} \vee
Q_{i_2}$. Suppose now that the instants $t_{e,1}$ and $t'_{e,1}$
have been chosen and that the vertex $\alpha$ (resp. $\beta$) occurs
at time $t_{e,1}$ (resp. $t'_{e,1}$) (the other case can be
handled similarly). We claim that the number of choices for
the instant $t'_{e,2}$ is at most of $2(\nu _N + T_N)$ instead of $s_N$. We recall
that the quantities $\nu _N$ and $T_N$ are such that, given an
arbitrary vertex $v \not = 1$ in $P''$, there are at most $\nu _N+T_N $ up
steps having $v$ for endpoint. The announced bound readily
follows since the edge started at $t'_{e,2}$ has $\alpha$ or $\beta$ (already determined by the choices of $t_{e,1}$ and $t'_{e,1}$) as left endpoint. Obviously, the reasoning also applies to the $n_{e}-2$ remaining glued subpaths
$Q_{i_1} \vee Q_{i_2}$ having $e$ as associated erased edge.\\
We now consider the weight of the path $P$ with respect to that of $P''$.
The weight of the erased edge $e$ must multiply that of $P''$ : we use
the fact that  
$$\mbE |X_e|^{2n_e}\leq \sqrt{\mbE |X_e|^4}\sqrt{\mbE |X_e|^{4n-4}}\leq (Cn_e)^{n_e-1}.$$
Combining all the preceding, we deduce that $Z_5^0$ is of order 
\begin{eqnarray*}&&(\ref{pourZ5}) \times \sum_{n\geq 1}\frac{1}{n!}\left (\frac{4C(\nu_N+T_N) n}{s_N}\right)^{n}
\leq \frac{C(\nu_N+T_N)}{N^{2/3}} \times (\ref{pourZ5}).
\end{eqnarray*}
$Z_5^0$ is thus negligible with respect to $Z_4$ (and so $Z_1$) since
$\nu_N+T_N<<\sqrt{s_N}$ in typical paths $P''$ (cf.
\cite{PecheWishart}, Section 3.2).

\paragraph{}We now estimate $\hat Z _5$.
Consider a path $P$ which contributes to $\hat Z_5$ and let $e$ be an erased edge in-between $P'$ and $P''$ which appears 4 times or more
in $P'$. We also denote by $n_e \geq 1$ the number
of times where $e$ is an erased edge in the gluing process. As in typical paths $P''$ each edge is passed twice, $e$ appears exactly $2(n_e+1)$ times in $P$. The previous reasoning works allowing $n_e \geq 1$. We then see that the sole case which remains to be considered is $n_e=1$
that is when $e$ occurs $4$ times and not more in $P$. In this case,
we determine the instants $t_{e,1}$ and $t'_{e,1}$ as in the estimate
of $Z_4$. This determines the edge $e=\begin{pmatrix}\alpha
  \\\beta\end{pmatrix}$. Now, the knowledge of $e$ decreases the
possible choices of the marked occurrence of $e$ in $P''$. More
precisely, one pays a cost of order
$(\nu_N+T_N)^2/s_N$ so that a marked occurrence of $\beta$ (for
instance) arises in $P''$ after an occurrence of $\alpha$ (see also
\cite{Si-So1}, p. 13).
Using the above it is not hard to deduce that $\hat Z _5<<Z_1.$
This finishes the proof of Proposition \ref{prop: estZ2'm not=s}. $\square$

\section{Higher moments} \label{sec: higher}
Let $K$ be a fixed integer. Let also $c_i, i=1, \ldots K,$ be some
positive real numbers. In this section, we compute moments of the
type
$\mbE \Bigl (\prod_{i=1}^K {\rm Tr} V_N^{s_N^{(i)}}\Bigr),$
where $(s_N^{(i)})$ are some sequences of integers such that
$\lim_{N \to \infty} {s_N^{(i)}}/{N^{1/2}}=c_i \text{ if $\pi_1>w_c$} \text{ and }$
$\lim_{N \to \infty} {s_N^{(i)}}/{N^{2/3}}=c_i\text{ if $\pi_1\leq w_c$}.$
Then we prove the following result.
Set 
$$
\tilde V_N^{(G)}=\frac{V_N^{(G)}}{u_+}\text{ if $\pi_1\leq w_c$} \: \text{ and } \: \tilde V_N^{(G)}=\frac{V_N^{(G)}}{\tau(\pi_1)}\text{ if $\pi_1> w_c$}.$$

\bp \label{Prop: variance}Under the assumptions of Theorems \ref{theo: unicomplex} and \ref{theo: unireal},
there exists a constant $C=C(K)>0$ such that $\mbE 
 \Bigl (\prod_{i=1}^K \tTr \,{\tilde V_N}^{s_N^{(i)}} \Bigr)\leq C $ and 
$$\mbE \left (\prod_{i=1}^K  \tTr \,{\tilde V_N}^{s_N^{(i)}} \right)=(1+o(1)) \mbE \left (\prod_{i=1}^K \tTr  \bigg (({\tilde V_N^G}) \,^{s_N^{(i)}} \bigg) \right).$$
In the case where $\pi_1\geq w_c$, the constant $C$ also depends on $\max_j \mbE |X_{1j}|^4$.
\ep
\paragraph{Proof of Proposition \ref{Prop: variance}:}
We consider the variance $ \mbE \left (\tTr V_N^{s_N}- \mbE \tTr V_N^{s_N}\right)^2$ only. Indeed the computations needed to consider higher moments follow from the same arguments combined with those developed in Section 5 of \cite{Si-So2}.
Proposition \ref{Prop: variance} can then be restated as follows.
\bl \label{Prop: variance'}
There exists $C>0$ such that
$\mbE \left (\tTr {\tilde V_N}^{s_N}- \mbE \tTr {\tilde V_N}^{s_N}\right)^2\leq C $ and one has \, $\mbE \left (\tTr {\tilde V_N}^{s_N}- \mbE \tTr {\tilde V_N}^{s_N}\right)^2=(1+o(1)) \mbE \left (\tTr {(\tilde V_N^G)}^{s_N}- \mbE \tTr {(\tilde V_N^G)}^{s_N}\right)^2.$
\el
\paragraph{Proof of Lemma \ref{Prop: variance'}:}Let us define $Y:=\Sigma^{1/2}X$. Then,
\begin{eqnarray*}
&& \!\!\!\!\!\!p^{2s_N}\mbE \left (\tTr V_N^{s_N}- \mbE \tTr V_N^{s_N}\right)^2\crcr
&&\!\!\!\!\!\!= \sum_{P_{(1)}, P_{(2)}}^* \mbE\left (\prod_{(i,j) \in P_{(1)}} \hat{Y}_{ij}\prod_{(i,j) \in P_{(2)}} \hat{Y}_{ij}\right )-\mbE \left(\prod_{(i,j) \in P_{(1)}} \hat{Y}_{ij} \right )\mbE\left (\prod_{(i,j) \in P_{(2)}} \hat{Y}_{ij}\right ). 
\end{eqnarray*} 
Here, given an edge $e = (i,j)\in P_{(1)}$ (this is similar for $P_{(2)}$), $\hat Y_{ij}$ stands for $Y_{ij}$ if e occurs at an odd
instant of $P_{(1)}$ and for $\overline{Y_{ji}}$ if it occurs at an
even instant. The starred sum bears on paths $P_{(1)}, P_{(2)}$ of length
$2s_N$ sharing at least one common edge $(i,j)$, $i\in [1, \ldots, N],
j\in [1, \ldots, p]$. This follows from the fact that the $Y_{ij}$'s are independent centered random variables.
We say that such paths are \emph{correlated paths}. The contribution to the variance from correlated paths without $1$-edges can be deduced from \cite{PecheWishart}. We thus focus on the pairs of
correlated paths with $1$-edges and assume without loss of generality
that $P_{(1)}$ has at least one $1$-edge.

\paragraph{}We first consider the case where both $P_{(1)}$ and $P_{(2)}$ have $1$-edges. We denote by $T_1$ (resp. $T_2$) the number of pairs of $1$-edges in $P_{(1)}$ (resp. $P_{(2)}$). We also set $s=T_1+T_2$. We build from $P_{(1)}$ (resp. $P_{(2)}$) $T_1$ (resp. $T_2$) subpaths $(P_i)_{1 \leq i \leq T_1}$ 
(resp. $(P_i)_{T_1+1 \leq i \leq s}$) starting and ending with a $1$-edge as in Section \ref{sec: counting}. In the following, we use the denomination ``1-subpath'' or simply ``subpath'' of
$P_{(1)}$ or $P_{(2)}$ to refer to some subpath $P_i$. By definition, the origin of $P_{(1)}$
(resp. $P_{(2)}$) occurs at some even instant in the subpath $P_1$
(resp. $P_{T_1+1}$). We concatenate the subpaths $(P_i)_{1 \leq i \leq s }$ in the order they appear which leads to an even path $P$ of length $4s_N$. 
\paragraph{Case 1:} The subpaths $P_i$ in
$P_{(1)}$ and those of $P_{(2)}$ share $1$-edges only. Then, as in Section \ref{sec: counting}, we define the $l \leq s$ clusters of $P$ and we apply the first gluing procedure yielding a path
$P'$ of length $2(2s_N - (s- l))$. We denote by $x'$ the trajectory
of $P'$ and by $m$ the number of returns to 0 of $x'$. As $P_{(1)}$ and $P_{(2)}$ are
correlated, one has that $l <s$. Here, we will also assume that $P'$
is fundamental that is $m=l$. Otherwise, this implies to perform the
second gluing procedure on $P'$ yielding a new path $P''$ but as we
assume that the paths $P_{(1)}$ and $P_{(2)}$ share $1$-edges only,
all the arguments we will give to determine $P_{(2)}$ from $P'$ are exactly the same when dealing with $P''$ (see below). Thus, focusing on $P'$, $s-m$ counts the number of pairs of $1$-edges that have
been erased through the first gluing process. For the sequel, it is convenient to denote by $2L_j, \, j=1, \ldots,m$ the length of the successive $m$ sub-Dyck paths of $x'$ ($\sum _j 2 L_j= 2(2s_N - (s- l))$).\\
To reconstruct $P_{(1)}$ and $P_{(2)}$ from $P'$, one has to
determine the $s-m$ instants of time where a $1$-edge has been erased and
reorder the subpaths thus defined. One also has to determine the
origins of $P_{(1)}$ and $P_{(2)}$. By construction, the origin of
$P_{(1)}$ occurs at some even instant in the first 1-subpath in $P'$. We call $t_e$ the first moment of time where a 1-subpath of
$P_{(2)}$ is glued to a 1-subpath of
$P_{(1)}$. We call $Q$ the latter subpath of $P_{(2)}$. One can note
that at time $t_e$, a $1$-edge which we call $e$ is erased. Last, we
let $t_f$ be the instant where $Q$ stops in $P'$. Two cases must be
considered now since $t_f$ can be an instant where  
a $1$-edge is erased or where $x'$ returns to 0.\\
\indent Assume first that $t_f$ is an instant where
a $1$-edge is erased. Assume that $t_f$ and all but $t_e$ of the $s-m-2$ other moments
of time where a $1$-edge is erased have been selected in $P'$. Assume
also that the corresponding $(s-1)$ $1$-subpaths have been reordered. There are $\binom{s_N}{s-m-1}$ possible choices for the $s-m-1$
instants of the erased edges and ${(s-1)!}/{m!}$ ways to reorder
the $1$-subpaths thus defined. Indeed, the $m$ subpaths beginning the
sub-Dyck paths of $x'$ arise in the same relative order in $ P'$
and in the concatenation $P$ (cf. the proof of Lemma \ref{Lemm:
  nbrpreimagesP'}). We call $\hat P$ the path obtained
after rearranging these $s-1$ subpaths. We now choose along $\hat P$ the instant $t_0$ defining the origin
of $P_{(2)}$ : this determines all the subpaths of $P_{(2)}$ except
$Q$. There are at most $2s_N$ choices for $t_0$. A crucial fact now is
that the knowledge of $t_0$ combined with that of $t_f$
determines the instant $t_e$ (in $\hat P $) since as $P_{(2)}$ is of
length $2s_N$, the length of $Q$ is then known. To obtain the full
path $P_{(2)}$ and the final concatenation $P$, it remains to insert
$Q$ in $\hat P $; there are at most $2s$ ways to do this (the factor
$2$ comes from the choice of the direction of reading $Q$ in $P$). Set
$ \tilde s_N=2s_N-1$ and $\hat{C}_4:= 1 + \max _v \mathbb E|X_{1v}|^4 /\sigma^2$. Combining the whole,
we get (for details, see the computations of $Z_2$ and $Z_3$ made in Section
\ref{subsec: casl=m}) that the contribution $Z_{v,1}^{(1)}$ to the
variance from correlated paths $(P_{(1)},P_{(2)})$ such that $t_f$ is
an instant where a $1$-edge is erased is at most (for some constant $C >0$) 
\begin{eqnarray}{\label{estZv11}}
&Z_{v,1}^{(1)} \leq & C \sigma ^{2\tilde s_N} \sum_{s_1=1}^{\tilde s_N}\sum_{s=1}^{\tilde s_N-s_1} 
\sum_{m=1}^{s-1} \sum_{k=1}^{\tilde s_N} \sum_{k_1\leq k+s-m} s_1 \frac{\gamma_N^{k}}{\gamma_N^{\tilde s_N}}  {\bf{N}}(s_1-1,
s_1-k_1)\crcr && {\bf{N}}(\tilde s_N-(s-m)-s_1,
k-k_1,m-1)\frac{\left (\frac{2\hat C_4ss_N}{N} \right)^{s'}}{s'!}
\pi_1^s  \frac{4s s_N \hat{C}_4}{N} {\label{estZv11bis}}\\
& & := \frac{4 C\hat{C}_4 s_N^{3/2}}{N} \sigma ^{2\tilde s_N} \times A_{\tilde s_N}
\end{eqnarray}
where we let $s':=s-m-1$. The factor ${\hat{C}_4}/{N}$ comes from the weight of the erased
1-edge $e$ (it can indeed be shown that $e$ can only occur at most
four times in typical paths $P$). In the case where $s_N =O(\sqrt{N})
$ and $\pi_1 >w_c$, we readily deduce that
$({\ref{estZv11}})/(\tau(\pi _1))^{2s_N}$ is bounded
(universality is discussed at the end of this
section). In the case where $\pi_1\leq w_c$ and 
$s_N=O(N^{2/3})$, it is a small computation, using the same arguments as in Lemma \ref{lem: coeffsn}, to check that 
\be A_{\tilde s_N}=
O(1)\frac{u_+^{\tilde s_N}}{\sqrt{\tilde s_N}}  \text{ if }\pi_1 <w_c \text{ and }
A_{\tilde s_N}=O(1)u_+^{\tilde s_N} \text{ if }\pi_1 =w_c.\label{AN}
\ee

\indent Assume now that the instant $t_f$ is such that $x'(t_f)=0$. We then fix $t_f$ by
choosing one such instant : this fixes some $1 \leq j \leq m$. 
Assume that all but $t_e$ of the $s-m-1$ other moments
of time where a $1$-edge is erased have been selected in $P'$. The knowledge of $t_f$ and of the $s-m-1$
selected instants determines the subpath $Q_0$ in $P'$ which still has to be
split into a subpath of $P_{(1)}$ and the first subpath, which we call
$Q$, of $P_{(2)}$ that is glued to one subpath of $P_{(1)}$. By construction, $Q_0$
is included in the sub-Dyck path of $x'$ ending at time $t_f$ so that the length of
$Q_0$ is not greater than $2L_j$. As before we reorder the $(s-1)$
1-subpaths thus defined to get the path $\hat P$. Now given $\hat P$, we claim that there are at most $8s_N$ different ways to choose $t_f$ and the instant $t_0$ defining (along $\hat P$) the origin of $P_{(2)}$. Indeed, in
the final concatenation (that is in $P$), the origin of $P_{(2)}$ is
encountered along the first subpath of $P_{(2)}$. So in $\hat P$,
$t_0$ is either in $Q$ or in a 1-subpath which begins in the
interval of time $[2s_N, 2s_N+2L_j]$. Denoting by $2l'$ the
length of the 1-path beginning in $[2s_N, 2s_N+2L_j]$ but which
does not finish in this interval (if it exists), there exists some $L'_j$ such that $2l' \leq 2 L'_j$. Hence the number of possible choices for $t_f$ and $t_0$ is at most 
$\sum_{j=1}^m 2L_j+2L'_j \leq 8s_N$ which is what we
wanted. We then readily conclude that the contribution $Z_{v,2}^{(1)}$ of such correlated paths $(P_{(1)},P_{(2)})$ behaves as $Z_{v,1}^{(1)}$ since it is at most 4 times the r.h.s of (\ref{estZv11bis}).
\paragraph{Case 2:} The paths $P_{(1)}$ and $P_{(2)}$ share
edges which are not $1$-edges. We denote by $Z_{v,1}^{(2)}$ the
contribution of such correlated paths $(P_{(1)},P_{(2)})$ to the
variance. Dealing with such a pair, we still apply the first gluing procedure on the concatenation $P$. 
If the path
$P'$ obtained in this way is such that all the $1$-edges arise
when the trajectory of $P'$ returns to the level 0, we can
finish the proof as before.\\
Otherwise, we apply the second gluing procedure defined in Section \ref{subsec: casl>m} getting a new path
$P''$ where each occurrence of the vertex 1 on the bottom line
corresponds to a marked instant or an instant where its trajectory $x''$ returns to 0. We denote by $m$ the number of returns to $0$ of $x''$ and by $2L_1, \ldots,  2L_m$ the length of the successive $m$ sub-Dyck paths of
$x''$. Given $P''$, we shall now reconstruct the paths $P_{(1)}$ and $P_{(2)}$. As before, we call $Q$ the first subpath of $P_{(2)}$ that is glued to one of $P_{(1)}$. We consider here the
case where $Q$ is glued using the second procedure which means that its gluing is associated to a marked occurrence of the vertex 1 in $P''$ (since the other case can be treated using the arguments developed in the previous case). We also denote by $t_e$ (resp. $t_f$) the instant of time where $Q$ begins (resp. ends) in $P''$. We assume that all the instants needed to define the gluing but that of $Q$ are chosen. We also assume that all the marked occurrences of $1$, except that associated to the gluing of $Q$, are known. All these instants define $(s-1)$ 1-subpaths which we reorder as before defining a new path called $\hat P$. Now if one also knows the instant $t_0$ defining the origin of $P_{(2)}$ in $\hat P$ and if $t_e$ is fixed, then the length of $Q$ is determined and there is no choice for $t_f$. In the sequel we set $l_{Q}=t_f-t_e$. We consider the case where the cluster containing the subpath $Q$ is
well separated from the others (not interlaced). The other case
follows from the same considerations. Then during the time interval
$[t_e, t_e + l_{Q}]$, the trajectory $x''$ of $P^{''}$ does not go
below the level $x''(t_e)$ and there is also a marked occurrence of
$1$ on the bottom line. Furthermore, by the definition of the second gluing procedure, $[t_e, t_e + l_{Q}]$ is included
in a sub-Dyck path of $P^{''}$. Assume that this is the $j$th sub-Dyck path, which is thus of length $2L_j$. Let also $k_j$ be the number of odd up steps in this sub-Dyck path. Denote by $N_{t_e}$ the total number of
possible choices for the instants $t_e$ and that of the marked
occurrence of 1 associated to the gluing of $Q$.  Let $\mathbb E_{L_j,k_j}$ denote the expectation with respect to the uniform distribution on the set $\chi_{L_j, k_j}$ of Dyck paths of length
$2L_j$ with $k_j$ odd up steps. Then there exists a constant $C>0$ independent of $N$, $k_j$ and $L_j$ such that (for typical $k_j$'s) 
\begin{equation}{\label{estENte}}
\mathbb E _{L_j,k_j} \left ({N_{t_e}} / {s_N^{3/2}} \right) \leq C.
\end{equation}
The above bound essentially follows from arguments close to those used in (\ref{estENte1}) and the estimation obtained in Section 2.5 in \cite{PecheWishart}. More precisely, setting $T_{0,n,k}:= \# \chi_{n,k}$ for any $n,k$, it is easy to show that 
$$ \mathbb E _{L_j,k_j} \left ( {N_{t_e}} \right) \leq \sum _{n,k'} \frac{4 n (L_j - n) T_{0,n,k'} T_{0,L_j - n, k_j-k'}}{T_{0,L_j,k_j}} \leq C s_N^{3/2}$$
where $n$ (resp. $L_j - n$) counts the number of possible choices of the instant of the marked occurrence of $1$ (resp. of the instant $t_e$) if the sub-Dyck path between $t_e$ and the first return to $x(t_e)$ followed by a down step is of length $2n$. The above estimate clearly holds if $L_j \leq s_N^{1/2}$ and if $L_j\geq s_N^{1/2}$, one can copy the arguments of Section 2.5 in \cite{PecheWishart}. 
We are now in position to estimate the contribution
$Z_{v,1}^{(2)}$. To this aim, we denote by ${Z}_4'$ the contribution
$Z_4$ of Section \ref{subsec: casl>m} corresponding to even paths of
length $4s_N$ instead of $2s_N$. Apart from $N_{t_e}$, one
needs to multiply the contribution of $P''$ by a factor of the order ${16 \sigma ^2 ss_N}/{N^2}$. Indeed, the number of ways to determine the sub-Dyck path of $P^{''}$ where $[t_e, t_e+l_Q]$ is
included and the origin of $P_{(2)}$ in $\hat P ''$ may be
controlled as before by a factor $\sum_{j=1}^{m} 2L_j + 2L_j' \leq 8
s_N$. Besides, there are at most $2s$ ways to insert $Q$ in $\hat P$
and choose its orientation. Last, due to the edge $e$ erased at time
$t_e$ in between $P$ and $P''$ (it can be shown that $e$ does not
occur in typical paths $P''$ and occurs twice in typical paths $P$) and due to the marked occurrence of $1$ associated to $t_e$ and $t_f$, the weight of the path has to be multiplied by a factor $\sigma^2/ p \times 1/N$. Hence, inserting the factor $\sigma^2ss_NN_{t_e}/(pN)$ in the computations of ${Z}_4'$ and using $(\ref{estENte})$ and
$(\ref{AN})$, leads to 
$$Z_{v,1}^{(2)} \leq \frac{ C\sigma^2 s_N^{3/2}}{N} \times Z_{v,1}^{(1)},$$
for some positive constant $C$. One can then check that $Z_{v,1}^{(2)}=O(1){Z}'_4$. 
\paragraph{}To complete the analysis, we now investigate
the case where $P_{(2)}$ has no $1$-edge. In this case, we first apply the first gluing procedure to $P_{(1)}$, which leads to a path called $P'_{(1)}$. 
Then we use the second gluing procedure to ``insert'' $P_{(2)}$: we consider the first edge along $P'_{(1)}$ which is also encountered along $P_{(2)}$ and use it to glue $P_{(2)}$ by the construction procedure used in Section \ref{subsec: casl>m}. Last we use the second gluing procedure (if needed) to obtain a final path where all the $1$-edges arise at level $0$ of the associated trajectory or correspond to a marked occurrence of $1$. The procedure we use in this case can be compared to that of Case 2, provided the path $P_{(2)}$ is ``assimilated'' to a $1$-subpath.
The analysis performed in Case 2 can be copied up to minor modifications to show that the contribution to the variance of such correlated paths is of the order of $Z_{v,1}^{(2)}.$
\paragraph{}Combining all the preceding implies that the total contribution to the
variance $\text{Var}
\bigl(\text{ Tr}\tilde V_N^{2s_N} \bigr)$ from correlated paths $(P_{(1)}, P_{(2)})$ is
bounded. (\ref{AN}) also implies that in the case where $\pi_1<w_c$,
the contribution of paths with $1$-edges is negligible in the large $N$-limit. \\
To conclude to universality of the variance, we can use the fact that 
only the pairs of correlated paths with $1$-edges seen at most $4$
times and other edges passed exactly twice contribute in a non
negligible way to (\ref{estZv11}). Thus universality of the variance (and higher moments) can be deduced from universality of the expectation of traces of $V_N^{2s_N}$.
This finishes the proof of Proposition \ref{Prop: variance}. $\square$

\section{More than one eigenvalue greater than $1$ \label{Sec: r>1}}
In this section we consider more general spiked sample covariance
matrices $(V_N)$ given by $(\ref{def:
VN})$ with a spiked covariance matrix $\Sigma =\text{diag}(\pi_1,\pi_2, \ldots, \pi_r,1, \ldots,1)$
where $r \geq 2$ is some fixed integer independent of $p$ and $N$ and
$\pi_1 \geq \pi_2 \geq \cdots \geq \pi_r >1$ are given real numbers
independent of $p$ and $N$ also.\\
We shall explain the main modifications to be made in
the previous analysis in order to prove Theorems \ref{theo: unicomplex}
and \ref{theo: unireal} in this more complex case. As in the case where $r=1$ (recall Section \ref{Subsec: Core}
which includes the case where $r \geq 1$), one has to prove
boundedness and universality of moments (of any fixed order) of traces
of high powers of $V_N$. We here restrict ourselves to the study of the
expectation. Universality of moments of higher order of traces of $V_N$ then follows
from the same arguments as in the case where $r=1$ (see Section
\ref{sec: higher}). 

As before, $(s_N)$ denotes a sequence of integers that may grow to
infinity. In order to examine the contribution from paths $P$ to the
expectation $\mbE (\tTr V_N ^{s_N})$, one has to consider the number
of times each of the vertices $1, \ldots, r$ occurs on the bottom line
of $P$. To fix the idea of the analysis, we consider the case where $r=2$. The general case then follows from a straightforward extension of the arguments used when $r=2$.

\paragraph{}Let then a path $P$ of length $2s_N$ contributing to $\mbE (\tTr V_N ^{s_N})$ be given.
We assume that $P$ has $T_1$ (resp. $T_2$) pairs of $1$-edges (resp. $2$-edges) with $T_1+T_2 \geq 1$. We set $s:=
T_1+T_2$. To deal with such a path, we define a glued path $P'$. The
gluing procedure (leading to $P'$) defined in Section \ref{sec: counting} when $P$ has
only $1$-edges (or only $2$-edges) is modified in the following way
when $T_1 T_2>0$.
We first identify the instants $t_1<t_2<\cdots<t_s$ where the first edge of pairs of $1$-edges
or $2$-edges occur in the path. We call $e_i^{(l)}$ (resp. $e_i^{(r)}$) the left (resp. right) edge of these $s$ pairs of edges. Then, for
$i\geq 2$, we define the subpath $P_i$ as the subpath starting with
$e_{i-1}^{(l)}$ and ending at $e_{i}^{(r)}$. As before $P_1$ is the path starting
at $e_{s}^{(r)}$ and ending at $e_{1}^{(l)}$ (we concatenate the end and beginning
of $P$). Two subpaths $P_i$ and $P_{i'}$ are now said to belong to the same ``connected component'' if they share a $1$-edge or a $2$-edge. We denote by $l$ ($l \leq s$) the number of such connected components.
Consider the first connected component and denote by $l_1$ its cardinality. We claim 
(since each of the $1$- or $2$-edges occurs an even number of times, see Subsection \ref{subsec: glu}) that there exists a way to glue the $l_1$ subpaths in order to form a path satisfying the following conditions: it starts and ends with the same $1$-edge or $2$-edge and has no other $1$-edge or $2$-edge. 
We do the same for the other components in such a way to define $l$ paths (corresponding to each connected component) which have pairwise distinct first edges and appear in the same relative order as in the initial path $P$.
We denote by $Q_j, j=1, \ldots, l$ the successive paths derived from the gluing process (in case $r=1$, we denoted them by $P_j^g$).
To each path $Q_j$, we associate its connected component (also called cluster) $\mathcal S _j, j=1, \ldots, l$, which is the set of initial subpaths that have been glued to form $Q_j$. We then obtain a ``path'' $P'$ of length
$2(s_N-(s-l))$ with origin $1$ or $2$ and having $m$
returns to the level $0$, for some $m\leq s$. Note that (as $T_1 T_2>0$) $P'$ is not a path in the usual sense,
since one might switch from vertex $1$ to $2$ at any instant where one switches from one cluster to another.
Nevertheless each cluster (or subpath $Q_j$) starts and ends with the same vertex.
We start with the following important remark. Assume that the clusters
$\mathcal S _j$ and $\mathcal S _{j+1}$ do not have the same origin
and that, for instance, $\mathcal S _{j+1}$ starts with a $2$ (the reverse case is similar). This necessarily implies that some subpaths starting or ending with a $2$ have been glued in some preceding clusters. In other words, if we denote by $K$ the number of times one switches the origin of successive clusters, one has that
$ K \leq s-l .$

\paragraph{} We first assume that $l=m$ that is the returns of the trajectory associated to $P'$ to
the level $0$ define the $l=m$ clusters. Here we show that the contribution of paths for which $T_2>0$ is negligible if $\pi_2<\pi_1$. Their contribution is of the same order as that of paths with only $1$-edges (and only $2$-edges) in the case where $\pi_1=\pi_2.$

One of the main points in the analysis is to estimate the number of preimages $P$ of a glued
path $P'$, that is to establish the counterpart of 
Lemma \ref{Lemm:  nbrpreimagesP'}. The number of ways to determine the set $\mathcal T$ of the $s-m$ moments of time where some $1$- or $2$-edge has been erased is at most 
$\binom{s_N}{s-m}$ as before. Yet the number of ways to reorder the subpaths $P_i$ thus defined is much smaller than
in the case where $r=1$. When $r=1$, we used (recall the proof of
Lemma \ref{Lemm: nbrpreimagesP'}) the upper bound $\frac{s!}{m!}\leq
s^{s-m}$.  When $r=2$, there are some constraints on the way to reorder the subpaths $P_i$: they must be reordered in such a way to form a path. Indeed a subpath starting with a $1$-edge (resp. $2$-edge) cannot follow a subpath ending with a $2$-edge (resp. $1$-edge). 
When $r=2$ (and $T_1 T_2>0$) and assuming that the origin of $P'$ is chosen, the maximal number of ways to reorder the subpaths $P_i$
if one does not take these constraints into account is bounded by
$s^{s-m}$: for each cluster $\mathcal S _j$, it is enough 
to indicate the number of ``slots'' between the first subpath and each of the subpaths of $\mathcal S _j$. 
Let us call $R$ the number of ways to reorder the $P_i$'s in an admissible way now. 
Then if $T_1 T_2 >0$, one has that 
\begin{equation}{\label{BoundR}}
R \leq 8s^{s-m-1}.
\end{equation} 
To prove this, we need a few notations. We call $x_1$ (resp. $x_2$)
the number of subpaths $P_i$ starting and ending with a $1$-edge (resp. with a $2$-edge). And $2x_3:=s-x_1-x_2$
denotes the number of paths with both a $1$-edge and a $2$-edge. It will be convenient to call these paths respectively 
$1$-paths, $2$-paths or $12$-paths.
We here consider the set of paths for which $T_1 \geq T_2$ and which are obtained from an admissible configuration of the $P_i$'s.
Assume that $x_2\not=0.$
If $x_1\geq x_3$, consider all the configurations obtained by
permuting one of the $x_1$ $1$-subpaths with one of the $x_2$
$2$-subpaths. Then distinct admissible configurations lead to distinct non admissible configurations. Similarly, if $x_3\geq x_1$, we consider all the configurations obtained by permuting one of the $x_3$ $12$-subpaths with one of the $x_2$ $2$-subpaths.
If now $x_2=0$, we consider all the configurations obtained by permuting one of the $x_3$ $12$-subpaths with one of the $x_1$ $1$-subpaths.
In all these cases, the number of permutations is at least $ s/8$ which leads to $(\ref{BoundR})$.

\paragraph{}From now on, we assume that there exists a real number $c>0$ such that 
$\lim_{N } {s_N}/ {\sqrt N}=c \text{ if $\pi_1>w_c$}$ and $\lim_{N} {s_N}/{ N^{2/3}}=c \text{ if $\pi_1\leq w_c.$}$\\
In the following, we focus on the estimation of the
  contribution from paths $P$ having $1$-edges and
$2$-edges passed only twice. As in the case $r=1$ (and calling on
\cite{PecheWishart}) it is not hard to see that, amongst the associated glued
paths $P'$, the typical ones have edges passed at most twice.\\
Thus, the contribution of such paths $P$ with $T_1$ pairs of $1$-edges and $T_2$ pairs of $2$-edges ($T_1\geq T_2>0$) can be bounded from above by
\begin{eqnarray}
&& C \sigma^{2s_N} \sum_{s_1=1}^{s_N}\sum_{s=0}^{s_N-s_1}
\sum_{m=1}^{s} \sum_{K=1}^{m} 1_{K\leq s-m}\binom{m}{K}\sum_{ k, k_1}\crcr
&&s_1 {\bf{N}}(s_1-1, s_1-k_1)  {\bf{N}}(s_N-s_1-(s-m), k-k_1,m-1) \crcr
&& \left ( \frac{2 \gamma _N ^{-2}}{N} \right ) ^{s-m} \binom{s_N-(s-m)-1}{s-m}  \left (Cs \right)^{s-m-1}\gamma_N^{k+(s-m)-s_N}\pi_1^{T_1}\pi_2^{T_2},
\end{eqnarray}
where the extra factor $\binom{m}{K}$ comes from the fact that we have to distribute the $m$ starting points of clusters into those starting with $1$'s and $2$'s (and $C$ is a positive constant whose value may vary in the following).
The contribution of paths for which $T_2\geq T_1>0$ can be analyzed in a similar way. One simply interchanges the role of $x_1$ and $x_2$ in the previous reasoning.

As $\sum_{K=1}^{m}1_{K\leq s-m}\binom{m}{K}\leq 2^m$, it is clear that
the contribution of the paths $P$ such that $m\leq 100(s-m)$ (100 is an arbitrarily large constant here) yields a contribution which is at most in the order of 
\begin{eqnarray}&& C \sigma^{2s_N} \sum_{s_1=1}^{s_N}
  \sum_{s=0}^{s_N-s_1}  \sum_{m=1}^{\frac{100}{101}s} \sum_{K=1}^{m}
  1_{K\leq s-m}  \binom{m}{K} \sum_{ k, k_1}\crcr
&&s_1 {\bf{N}}(s_1-1, s_1-k_1) {\bf{N}}(s_N-s_1-(s-m), k-k_1,m-1) \crcr
&&\left ( \frac{2 \gamma _N ^{-2}}{N} \right ) ^{s-m} \binom{s_N-(s-m)-1}{s-m}  \left (Cs \right)^{s-m}\gamma_N^{k+(s-m)-s_N}\pi_1^{s} 
\frac{1}{s}
\sum_{T_2\leq s} \left( \frac{\pi_2}{\pi_1}\right)^{T_2}\crcr
&&=\begin{cases}O(\frac{s_N}{N})Z_1 & \text{ if }\pi_2<\pi_1,\\
    O(1) Z_1 &\text{ if }\pi_2=\pi_1.
   \end{cases}
\end{eqnarray}
where $Z_1$ is given by (\ref{est: s=m}).\\
There now remains to estimate the contribution of paths $P$ for which $m>100(s-m)$.
To this aim, we need to refine our preceding reasoning. Assume
that the $s-m$ moments of time of the set $\mathcal T$ as well as the $K$
instants where one switches the origin of the clusters have been selected. Assume also for ease
that the origin of the path $P'$ is $1$ and denote by $m_1$ (resp. $m_2$) the number of $1$-paths (resp. $2$-paths) starting a cluster. Last set $m_3=m-m_1-m_2.$
To reorder the $P_i$'s, we first reorder the $2x_3$ $12$-paths. There are $\frac{(2x_3)!}{m_3!}$ ways to do so. Then we determine the number of 
$1$-paths and $2$-paths to be inserted in-between the $12$-paths and reorder them.
There are at most 
$\frac{x_1!}{m_1!}\frac{x_2!}{m_2!} 2^{s-m}$ ways to do so (the $2^{s-m}$ is
due to the possible choice of the direction of reading each of the
$1$- and $2$-paths).

Thus the contribution of paths $P$ for which $m\geq 100(s-m)$ can be bounded from above by 
\begin{eqnarray}&& C \sigma^{2s_N} \sum_{s_1=1}^{s_N} \sum_{s=0}^{s_N-s_1}\sum_{m\geq \frac{100}{101}s} \sum_{K=1}^{m} 1_{K\leq s-m}  \binom{m}{K} \sum_{ k,k_1}\crcr
&&s_1 {\bf N} (s_1-1, s_1-k_1) {\bf N} (s_N-s_1-(s-m), k-k_1,m-1) \left ( \frac{2 \gamma _N ^{-1}}{N} \right ) ^{s-m} \crcr
&& \binom{s_N-(s-m)-1}{s-m}  \frac{1}{m_1!m_2!m_3!}\left (Cs
\right)^{s-m} \gamma_N^{k-s_N} \pi_1^{s}\sum_{T_2\leq s} \left( \frac{\pi_2}{\pi_1}\right)^{T_2}.
\end{eqnarray}
Using the fact that $\frac{1}{m_1!m_2!m_3!}\leq \frac{1}{(m/3)!}\frac{3\times 101/100}{s}$ and that $\sum_K \binom{m}{K}\leq 8^{m/3},$
it is clear that the contribution of paths for which $s\geq \sqrt{s_N}$ is negligible.
The contribution of paths for which $s\leq \sqrt{s_N}$ can be analyzed as follows.
If $\pi_1 >w_c$, their contribution is of order $u_+^{s_N}\pi_1^{\sqrt{s_N}}<<\tau(\pi_1)^{s_N}$ and is thus negligible.
If $\pi_1\leq w_c$ it is not hard to see that their contribution is at
most of order of $Z_1$ (and thus negligible if $\pi_1<w_c$). 

The contribution from paths $P$ (such that $l=m$) having $1$-edges and $2$-edges
possibly read more than twice can be examined by refining the above
analysis thanks to arguments already used in Section 
{\ref{sec: expect}} for the investigations of $Z_3$. We skip the detail. Thus, one can show that paths $P$ such that $l=m$ satisfy:\\
(a) if $\pi_1=\pi_2 \geq w_c$, the typical paths $P$ have $1$-edges and $2$-edges seen at most $4$ times  and no other edge seen more than twice;\\
(b) if $\pi_1>\pi_2$, the typical paths have no $2$-edge;\\
(c) if $\pi_1<w_c$, the typical paths have neither $1$-edges nor $2$-edges.\\
Last the contribution from paths $P$ such that $l>m$ that is when some occurrences of $1$- or $2$-edges in $P'$ arise at some positive level can be analyzed using the same arguments as in Section \ref{subsec: casl>m} and arguments as above. We then deduce that when $r=2$, the typical paths contributing to $\mbE
(\tTr V_N ^{s_N})$ satisfy the three preceding conditions (a) to (c). Combining all the preceding justifies the universality of the expectation $\mbE
(\tTr V_N ^{s_N})$.


\addcontentsline{toc}{chapter}{Bibliographer}
\markboth{BIBLIOGRAPHIE}{BIBLIOGRAPHIE}
\begin{thebibliography}{50}
\bibitem{PlerousGRAGS}\,
\textsc{Amaral, L.}, \textsc{Gopikrishnan, P.}, \textsc{Guhr, T.}, \textsc{Plerous, V.}, \textsc{Rosenow, B.} and \textsc{Stanley, H.},
\newblock Random matrix approach to cross correlations in financial data.
\newblock \textit {Phys. Rev. E }\textbf{65} no. 6, 66--126 (2002).


\bibitem{Bai}\,  \textsc{Bai, Z.D.}, \newblock {Methodologies in spectral analysis of large-dimensional random matrices, a review.}
\newblock \textit {Statist. Sinica} \textbf{9} no. 3, 611--677 (1999).

\bibitem{BaiYao}\,
\textsc{Bai, Z.} and \textsc{Yao, J.}, Central limit theorems for eigenvalues in a spiked population model. {\it Ann.
  Inst. H. Poincar\'e} \textbf{44} no. 3, 447--474 (2008).

\bibitem{Baik} \, \textsc{Baik, J.}, \newblock{Painlev\'e formulas of the limiting distributions for nonnull
              complex sample covariance matrices.} \textit{Duke Math. J.}
              \textbf{133} no. 2, 205--235 (2006).

\bibitem {BaikGBAPeche} \, \textsc{Baik, J.}, \textsc{Ben Arous, G.} and \textsc{P\'ech\'e, S.}, {Phase
    transition of the largest eigenvalue for non-null complex sample
    covariance matrices.} \textit{ Ann. Probab. }\textbf{33} no. 5, 1643--1697 (2005).

\bibitem{BaikSilverstein} \, \textsc{Baik, J.} and \textsc{Silverstein, J.}, {Eigenvalues of large sample covariance matrices of spiked population models.}
\newblock \textit {Journ. of Mult. Anal.} \textbf{97}, 1382--1408 (2006).

\bibitem{Bouchaud} \textsc{Biroli, G., Bouchaud, J.P.} and \textsc{Potters, M.}, {On
    the top eigenvalue of heavy-tailed random matrices.} \textit {Europhysics Letters} \textbf{78,} 10001 (2007).

\bibitem{LalouxCPB}
\textsc{Bouchaud, J.}, \textsc{Cizeau, P.}, \textsc{Laloux, L.} and \textsc{Potters, M.}, \newblock Random matrix theory and financial correlations.
\newblock \textit {Intern. J. Theor. Appl. Finance} \textbf{ 3} no. 3, 391--397 (2000).

\bibitem{Chen}\,
\textsc{Chen, W.}, \textsc{Yan, S.} and \textsc{Yang, L.}, Identities from Weighted $2$ Motzkin paths. Available at www.billchen.org/publications/identit/identit.pdf.

\bibitem{NEKlarginterest}
\textsc{El Karoui, N.}, \newblock Recent results about the largest eigenvalue of random covariance matrices and statistical application.
\newblock \textit {Acta Phys. Polon. B} \textbf{ 36} no. 9, 2681--2697 (2005).


\bibitem{PecheFeral}\,  \textsc{F\'eral, D.} and \textsc{P\'ech\'e, S.}, {The largest eigenvalue of rank one deformation of large Wigner
  matrices}. {\it Comm. Math. Phys.} {\bf{272}}, {185--228} (2007).

\bibitem{HoyleR}\,
\textsc{Hoyle, D.} and \textsc{Rattray, M.}, \newblock Limiting form of the sample covariance eigenspectrum in {PCA} and kernel {PCA}.
\newblock \textit {Advances in Neural Information Processing Systems NIPS 16} (2003).



\bibitem{Johansson}\, \textsc{Johansson, K.}, Shape fluctuations and random matrices.
\newblock \textit {Comm. Math. Phys.} \textbf{209}, 437--476  (2000).

\bibitem{Johnstone} \, \textsc{Johnstone, I. M.}, On the distribution of the largest {P}rincipal {C}omponent.
\newblock \textit {Ann. Statist.}
\textbf{29}, 295--327 (2001).

\bibitem{MalevergneS}
\textsc{Malevergne, Y.} and \textsc{Sornette, D.}, \newblock Collective origin of the coexistence of apparent {RMT} noise and factors in large sample correlation matrices.
\newblock \textit {Physica A} \textbf{331} no. 3-4, 660--668 (2004).

\bibitem{MP}\, \textsc{Marchenko, V.A.} and \textsc{Pastur, L.A.}, Distribution of eigenvalues for some sets of random matrices.
\newblock \textit {Math. USSR-Sbornik} \textbf{1}, 457--486 (1967).

\bibitem{Onatski}
\textsc{Onatski, A.}, The Tracy-Widom limit for the largest eigenvalues of singular complex Wishart matrices. 
\newblock \textit{ Ann. Appl. Probab.} \textbf{18}  no. 2, 470--490 (2008).

\bibitem{Debashis}\,
\textsc{ Paul, D.}, Asymptotics of the leading sample eigenvalues for a spiked covariance model. \textit{Stat. Sinica}
\textbf{17}, 1617--1642 (2007).

\bibitem{Paterson}\,
\textsc{Patterson, N.}, \textsc{Price, A.L.} and \textsc{Reich, D.}, \newblock Population structure and eigenanalysis.
\newblock \textit {PLoS Genet 2(12): e190 DOI: 10.1371/journal.pgen.0020190} (2006).

\bibitem{PecheWishart} \textsc{P\'ech\'e, S.},
\newblock Universality results for largest eigenvalues of some sample covariance matrix ensembles.  
\newblock {Accepted for publication in \textit{Prob. Th. Relat. Fields}. arXiv:math/0705.1701.} (2008).

\bibitem{PecheHDR}\textsc{P\'ech\'e, S.}, \newblock{The edge of the spectrum of random matrices. }
\newblock Habilitation Thesis, Universit\'e Joseph Fourier Grenoble (2008). 

\bibitem{PecheSos1} \textsc{P\'ech\'e, S.} and \textsc{Soshnikov, A.},
\newblock Wigner random matrices with non symmetrically distributed entries.
\textit{Journ. Stat. Phys.} \textbf{129} no. 5-6, 857--884 (2007).

\bibitem{SearC}
\textsc{Sear, R.} and \textsc{Cuesta, J.}, \newblock Instabilities in complex mixtures with a large number of components.
\newblock \textit {Phys. Rev. Lett.} \textbf{91} no. 24, 245--701 (2003).

\bibitem {Si-So1} \, \textsc{Sinai, Y.} and \textsc{Soshnikov, A.}, Central limit
    theorem for traces of large random symmetric matrices with independent
    matrix elements. \newblock {\it Bol. Soc. Brasil. Mat. (N.S.)} \textbf{29}, 1--24 (1998).

\bibitem {Si-So2} \, \textsc{Sinai, Y.} and \textsc{Soshnikov, A.}, A refinement of
    Wigner's semicircle law in a neighborhood of the spectrum edge for random
    symmetric matrices. {\it Funct. Anal. Appl.} \textbf{32}, 114--131 (1998).

\bibitem {Sos} \, \textsc{Soshnikov, A.}, Universality at the edge of the
    spectrum in Wigner random matrices. {\it Comm. Math. Phys.} \textbf{207}, 697--733 (1999).

\bibitem{SosWish}\, \textsc{Soshnikov, A.}, A note on universality of the distribution of the largest eigenvalues in certain sample covariance matrices.
\newblock  \textit {J. Statist. Phys.} \textbf{108}, 1033--1056 (2002).

\bibitem{Telatar}
\textsc{Telatar, E.}, \newblock Capacity of multi-antenna {G}aussian channels.
\newblock \textit {Europ. Trans. Telecom.} \textbf{10} no. 6, 585--595 (1999).

\bibitem{TWAi} \, \textsc{Tracy, C.} and \textsc{Widom, H.}, 
\newblock Level-spacing distribution and Airy kernel.
\newblock \textit {Commun. Math. Phys.} \textbf{159}, 151--174 (1994).

\bibitem{TWAi2} \, \textsc{Tracy, C.} and \textsc{Widom, H.}, On orthogonal and symplectic matrix ensembles.
\newblock \textit {Commun. Math. Phys.} \textbf{177}, 727--754 (1996).

\bibitem{Yin} \, \textsc{Yin, Y. Q.}, \textsc{Bai, Z. D.} and
  \textsc{Krishnaiah, P. R.}, On the limit of the largest eigenvalue
  of the large-dimensional sample covariance matrix,
  \textit{Prob. Th. Relat. Fields} \textbf{78}, 509--521 (1988).

\bibitem {Wigner} \, \textsc{Wigner, E.}, {Characteristic vectors of bordered matrices with infinite dimensions.} \textit{ Ann. Math. }\textbf{62}, 548--564 (1955).
\end{thebibliography}
\end{document}